%% file: main.tex
\documentclass{aastex63}

\newcommand\mb{\mathbf}

\received{December 31, 2021}
\revised{February 1, 2022}
\accepted{February 7, 2022}
\submitjournal{ApJ}

\shorttitle{Sample article}
\shortauthors{Daldorff et al.}
\graphicspath{{./}{figures/}}

\begin{document}
\title{Impact of 3D Structure on Magnetic Reconnection}

\correspondingauthor{Lars K.~S.~Dalorff}
\email{}

\author[0000-0002-1198-5138]{Lars K.~S.~Daldorff}
\affiliation{Catholic University of America, Washington DC, USA}
\affiliation{Heliophysics Science Division \\ NASA Goddard Space Flight Center \\ 8800 Greenbelt Rd.\\ Greenbelt, MD 20771, USA}

\author{James E.~Leake}
\affiliation{Heliophysics Science Division \\ NASA Goddard Space Flight Center \\ 8800 Greenbelt Rd.\\ Greenbelt, MD 20771, USA}
\author{James A.~Klimchuk}
\affiliation{Heliophysics Science Division \\ NASA Goddard Space Flight Center \\ 8800 Greenbelt Rd.\\ Greenbelt, MD 20771, USA}

\begin{abstract}

Results from 2.5D and 3D studies of the onset and development of the tearing instability are presented, using high fidelity resistive MHD simulations. A limited parameter study of the strength of the reconnecting field (or shear angle) was performed. An initially simple 1D equilibrium was used, consisting of a modified force-free current sheet, with periodic boundary conditions in all directions. In all cases, the linear and non-linear evolution led to a primary current sheet between two large flux ropes. The global reconnection rate during this later stage was analyzed in all simulations. It was found that in 2.5D the primary current sheet fragmented due to plasmoids, and as expected, the global reconnection rate, calculated using multiple methods, increases with the strength of the reconnecting field due to a stronger Alfv\'{e}n speed. In 3D, the presence of interacting oblique modes of the tearing instability complicates the simple 2.5D picture, entangling the magnetic field of the  inflow and introducing a negative effect on the reconnection rate. The two competing effects of stronger Alfv\'{e}n speed and entangling, which both increase with the strength of the reconnecting field, resulted in a decrease in the reconnection rate with increasing reconnecting field. For all simulations, the 3D rates were less than in 2.5D, but suggest that as one goes to weak reconnecting field (or strong guide field), the system becomes more 2.5D like and the 2.5D and 3D rates converge. These results have relevance to situations like nano-flare heating and flare current sheets in the corona. \\

\end{abstract}


\input{intro}

\input{numerics}

\input{results_initial}

\input{results_turbulence}

\input{results_rates}

\input{discussion}

\section{Acknowledgments}

This work was funded by NASA's Living With a Star (LWS) program and NASA's Internal Scientist Funding Model (ISFM).
The simulations for this work were run and analyzed on HECC systems, and the data and analysis tools to create the images in this paper are available upon request. The authors thank Y.Min Huang and Scott Baalrud for insightful discussion.

\bibliography{main}{}
\bibliographystyle{aasjournal}

\end{document}

%% file: intro.tex
\section{Introduction} \label{sec:intro}

 In the context of solar physics, the process of magnetic reconnection is believed to be the cause of transient phenomena such as solar flares, X-ray jets, and to play a significant role in coronal mass ejections (CMEs), as well as playing a crucial role in the heating of the atmosphere itself \citep[e.g.][]{Karpen_2012, Wyper_2017, Klimchuk_2015}. During magnetic reconnection, energy stored in magnetic fields is converted into kinetic and thermal energy of the plasma, and possibly also energetic particles, by breaking magnetic connectivity.  Since the magnetic reconnection rate scales with resistivity as  $\eta^{1/2}$ for steady state reconnection as described by Sweet and Parker \citep{Sweet_1958,Parker_1957}, and since resistivity is very low in the very conductive corona, a mechanism to achieve faster reconnection in the fluid regime has often been searched for.  Understanding the speed of magnetic reconnection in the solar context is important because it can affect the maximum temperature achieved in nanoflares and flares and the speed of CMEs.

A complete study of magnetic reconnection in the solar corona would cover from large scales (e.g., Mm), which dictate the build up and storage of the magnetic energy, down to the kinetic scales, where breaking of magnetic fieldlines and particle acceleration occurs. This is not practically possible at present. Fortuitously, recent work has highlighted that the onset of coronal magnetic reconnection can be studied in the fluid regime, where the flux-breaking is modeled by collisional resistivity. {Reconnection generally begins with the tearing instability \citep{1963PhFl....6..459F}.} While the fast breakup of Sweet-Parker scaled current sheets due to tearing (also called the  plasmoid instablity in this regime) had been well studied \citep[e.g.,][]{2002plap.book.....T,Loureiro_2007,Huang_2013}, more recent work has demonstrated that a thinning current sheet will exhibit fast tearing (also called \textit{ideal tearing}) at scales much larger than the Sweet-Parker width \citep{Pucci_2014,Del_Zanna_2016,Pucci_2017,Tenerani_2016}, so it can never reach the small widths predicted by the Sweet-Parker scaling without first fragmenting. The scales at which current sheets fragment, given coronal values of resistivity, may therefore be 100 times larger than kinetic scales, and at the very least, the onset of the current sheet breakup occurs on fluid scales. Hence, a fluid treatment of the problem can still be useful to understand the evolution, although heating and particle acceleration would require the treatment of the kinetic physics also.

The majority of the work performed on the tearing of current sheets has been conducted in 2.5D (where there are only two independent spatial dimensions, but all vector components are evolved), where it appears that the resultant reconnection rate can become independent of the resistivity \citep{Huang_2010, Uzdensky_2010,  Loureiro_2012}, but it is not clear whether this occurs in 3D. Based on this, and the discussion above, it should be clear that studying 3D tearing in coronal current sheets is an important area of study to understand the role of magnetic reconnection in coronal physics. 

\citep{Huang_2016} previously looked at reconnection in a current sheet, focusing on the effect of the 3D (oblique) modes of the system. 
In our study reported here, the effect of these oblique modes  on the effective (and mainly 2D) plasmoid-mediated reconnection in 3D current sheets is investigated. To do so, a parameter study of 2.5D and 3D simulations is performed by varying the initial reconnecting magnetic field strength (or rotation of the magnetic field in the current sheet) as this parameter affects the growth rate of the oblique modes. 

\S \ref{sec:numerics} summarizes the numerical setup, and details of the parameter study. \S \ref{sec:results} presents the results, first focusing on the initial development of the starting current sheet (\S\ref{sec:results_initial}), then focusing on the effect of the reconnecting field strength on the reconnection in 2D and 3D (\S \ref{sec:results_turbulence}) and then focusing on various measures of the normalized reconnection rate (\S \ref{sec:results_rates}). \S \ref{sec:discussion} summarizes and contextualizes the results.

%% file: numerics.tex
\section{Numerical Model and Setup} 
\label{sec:numerics}

The 3D simulations analyzed here are exactly those presented in \citet{2020ApJ...891...62L}, but the details of the numerics and initial conditions are included here for the reader's convenience.

The following visco-resistive MHD equations are solved, written here in Lagrangian form:

\begin{eqnarray}
\frac{D\rho}{Dt} & = & -\rho\nabla.\mathbf{V}, \\
\frac{D\mathbf{V}}{Dt} & = & -\frac{1}{\rho}\left[\nabla P
+ \mathbf{J}\times\mathbf{B} + \mb{F}_{shock}\right],\\
\frac{D\mathbf{B}}{Dt} & = & (\mathbf{B}.\nabla)\mathbf{V}
- \mathbf{B}(\nabla .\mathbf{V}) - \nabla \times (\eta\mathbf{J}), \\
\frac{D\epsilon}{Dt} & = & \frac{1}{\rho}\left[-P\nabla .\mathbf{V}
 + H_{visc} \right],
\label{eqn:energy_MHD}
\end{eqnarray}
where $\rho$ is the mass density, $\mathbf{v}$ the velocity, $\mathbf{B}$ the magnetic field, and $\epsilon$ the internal specific energy density. The current density is given by $\mathbf{J}=\nabla\times\mathbf{B}/\mu_{0}$, where $\mu_{0}$ is the permeability of free space, and $\eta$ is the resistivity. The gas pressure, $P$, and the specific internal energy density, $\epsilon$, can be written as
\begin{eqnarray}
P & = & \rho k_{B}T/\mu_{m}, \\
\epsilon & =  & \frac{k_{B}T}{\mu_{m}(\gamma-1)}
\label{eqn:eos}
\end{eqnarray}
 respectively, where $k_{B}$ is Boltzmann's constant and  $\gamma$ is 5/3. The reduced mass $\mu_{m}$, for this fully ionized hydrogen plasma, is given by $\mu_{m}=m_{p}/2$ where $m_{p}$ is the mass of a proton. The reader should note that the standard Ohmic heating term $\eta J^2$ has been removed from Eqn. (\ref{eqn:energy_MHD}), which is discussed below. 

The equations are solved using a Lagrangian-Remap (LaRe) approach \citep{Arber_2001}. 
The shock viscosity $\mathbf{F}_{shock}$ in the momentum equation is finite at discontinuities but zero for smooth flows, and the shock jump conditions are satisfied with an appropriate choice of shock viscosity \citep{Caramana_1998,Arber_2001}. This allows heating due to shocks to be captured as a viscous heating $H_{visc}$.

The equations above are non-dimensionalized by dividing each variable ($C$) by its normalizing value ($C_{0}$). The set of equations requires a choice of three normalizing values. The following values are chosen for the length, $L_{0}=10^{6}~ \textrm{m}$,  density, $\rho_{0}=1.5\times10^{-13} ~ \textrm{kg}/\textrm{m}^{3}$, and magnetic field, $B_{0}=10^{-3} ~ \textrm{T}$ ($10 ~ \textrm{G})$. This leads to a normalizing velocity of $V_{0} = 2.3\times10^{6} \textrm{m}/\textrm{s}$. 

 The  uniform background density and temperature are $1.67\times10^{-13} ~ \textrm{kg}/\textrm{m}^{3}$ and $10^{6} ~ \textrm{K}$. The magnetic field is a modified force-free Harris current sheet with a guide field, 
\begin{eqnarray}
B_x(y) & = & B_{x,0}\tanh{(\frac{y}{a})}\cos{(\pi \frac{y}{L_{y}/2})}, ~ \textrm{for} ~ |y| \le L_{y}/4 \\
& = & 0 , ~ \textrm{for} ~ |y| > L_{y}/4 \\ 
B_{z}(y) & = & \sqrt{(B_{z,0}^{2}-B_{x}(y)^{2})}
\end{eqnarray}

The cosine dependence on y reduces the shear component of the magnetic field ($B_{x}$) outside $\pm L_{y}/4$ and allows for the use of periodic boundary conditions in $y$ as well as the other two dimensions. It corresponds to a very broad return current that balances the thin current sheet we wish to study. For a given $B_{z,0}=10.6$ G, the parameter $B_{x,0}$ is varied to be [1.9,3.6,7.0] G. The shear angle,  which is the amount that the magnetic field rotates across the sheet, has a maximum value of $\theta = 2 \arctan(B_{x}(y)/B_{z}(y))$ in the domain, which, as shown in table \ref{table:IC}, is [20,40,80] degrees for the chosen $B_{x,0}$. 

The simulation grid extends [-50,50] Mm in $x$ and $y$, and [-25,25] Mm in $z$. There are (512,800,512) cells in the ($x,y,z$) directions. The $y$ domain has a non-constant resolution, where the grid size is a cubic function of the $y$ index. The minimum cell size in $y$ is 0.0125 Mm and the maximum is 0.34 Mm. The initial current sheet of half width $a=0.5 ~ \textrm{Mm}$ is well resolved as there are 75 cells across the full width in the $y$ direction.

Table \ref{table:IC} shows the 3D and 2.5D simulations analyzed in this paper. Note that the names of the simulations are different from what was used in \citet{2020ApJ...891...62L} to make discussion easier.

\begin{table}[h]
\centering
    \begin{tabular}{|c|c|c|c|c|c|c|c|}
    \hline
     Name & $B_{x,0}$ (G) &  $\theta$ & $V_{A,0} (10^6 ~ m/s)$ & $t_{A,0} (s)$ & $E_{free,0} (10^{18} ~ erg) $ & Name used in \\ 
     & & & & & & \citet{2020ApJ...891...62L} \\
    \hline
    Weak shear 3D: WS-3D & 1.90 & 20 & 0.44 & 228.51 & 1.93 & Sim 3 \\
    \hline
    Medium shear 3D: MS-3D & 3.60 & 40 & 0.83 & 120.60 & 6.94 & Sim 2 \\
     \hline
    Strong shear 3D: SS-3D & 7.00 & 80 & 1.61 & 62.02 & 26.23 & Sim 1 \\
       \hline
       \hline
    Weak shear 2D: WS-2D & 1.90 & 20 & 0.44 & 228.51 & $3.86\times10^8 \textrm{erg}/\textrm{cm}$ & -- \\
    \hline
    Medium shear 2D: MS-2D & 3.60 & 40 & 0.83 & 120.60 & $1.39\times10^9 \textrm{erg}/\textrm{cm}$ & -- \\
     \hline
    Strong shear 2D: SS-2D & 7.00 & 80 & 1.61 & 62.02 & $5.25\times10^9 \textrm{erg}/\textrm{cm}$ & -- \\  
    \hline
    \end{tabular}
\caption{Summary of initial conditions setup for all 6 simulations \label{table:IC}}
\end{table}

Given that the resistivity of the solar corona is so low, the numerical scheme will have a numerical resistivity much larger than the Sun's actual resistivity, so an  explicit value that is larger than the numerical value is used, $\eta=1.24\times10^{2} ~ \Omega.\textrm{m}$, or a diffusivity of $D=\eta/\mu_{0}=10^{8}  ~ \textrm{m}^2 \textrm{s}^{-1}$.

These choices of physical and closure parameters lead to a non-negligible Ohmic dissipation of magnetic energy, whereas in the solar corona it is negligible on all but very small scales. Therefore, while energy is lost from the magnetic field via Ohmic diffusion, the standard Ohmic $\eta J^{2}$ heating term has been removed from Eqn. \ref{eqn:energy_MHD} for the internal energy density $\epsilon$. As a consequence, the only form of  irreversable  heating in the simulations is caused by the shock heating described above. There is also heating and cooling form compression and expansion.

Numerical losses were calculated and found to be acceptable in the early stages of the evolution, where the initial current sheet is unstable to the 3D tearing mode \citet{2020ApJ...891...62L}. This is important as the locations of preferred tearing and the resultant structures heavily determine the later evolution, where, as will be shown, the coalescence of the tearing modes leads to a considerably thinner new current sheet. Due to the limitation of the numerical resolution in these simulations, this leads to increased numerical errors as this new current sheet is disrupted by new reconnection events (e.g., plasmoids). These new structures are thus not completely governed by the physical evolution, and care must be taken with conclusions based on the numerical results.

The linear stage of the 3D tearing instability is initialized using a superposition of weak Fourier modes, with mode numbers $(m,l,n)$,
which encompass the predicted allowed dominant parallel and oblique modes in this particular system, based on the linear theory below \citep{Baalrud_2012}:
 \begin{equation}
 V_{y}(x,y,z) = \sum_{m,l,n}{V_{0}}^{mln}{\sin{(k_{x}x-{\phi_{x}}^{mln})}\sin{(k_{y}y-{\phi_{y}}^{mln})}\sin{(k_{z}z-{\phi_{z}}^{mln})}}
 \end{equation}
where $k_{x} = 2\pi/\lambda_{x}=2\pi m/L_{x} ~( m=[1,n_m])$, $k_{y} = 2\pi/\lambda_{y} = 2\pi l /L_{y}~( l=[1,n_l])$, and $k_{z}=2\pi\lambda_{z} = 2\pi n/L_{z} ~(n=[0,n_n])$ where $(n_m,n_l,n_n)=(25,5,5)$. The phases ($\phi^{mln}$) and amplitudes ($V_{0}^{mln}$) for each mode are chosen randomly, with the amplitudes chosen randomly from a top-hat function in the range [-10, 10] m~s$^{-1}$. This results in an average initial velocity of 43 m~s$^{-1}$.

The 2.5D simulations are run using the same code, but by setting $L_z=0.01L_{0}$ and by using $n_z=2$, which with the periodic boundary conditions, and the removal of the velocity perturbation's dependence on $z$, creates a 2.5D version of the 3D simulations, and does not allow for variation in $z$ and so does not allow for the existence of oblique modes (see \S \ref{sec:results_initial}).

%% file: results_initial.tex
\section{Results}
\label{sec:results}
\subsection{Initial Evolution} 
\label{sec:results_initial}

\begin{figure}
    \centering
     \includegraphics[width=0.3\textwidth]{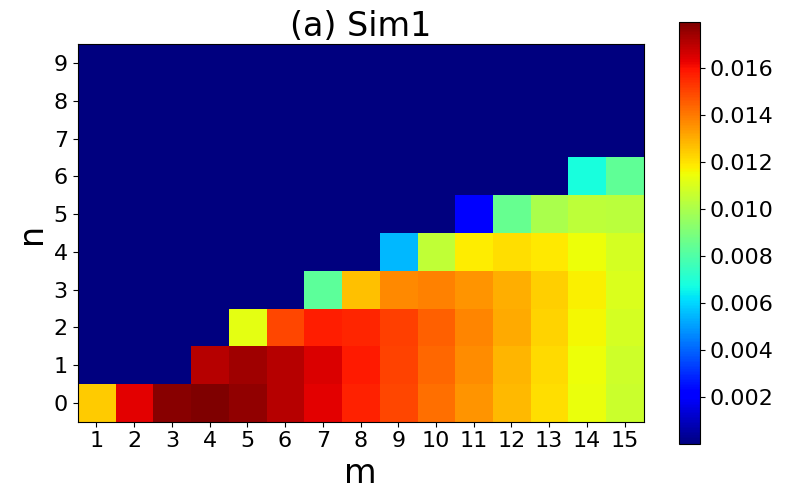}
    \includegraphics[width=0.3\textwidth]{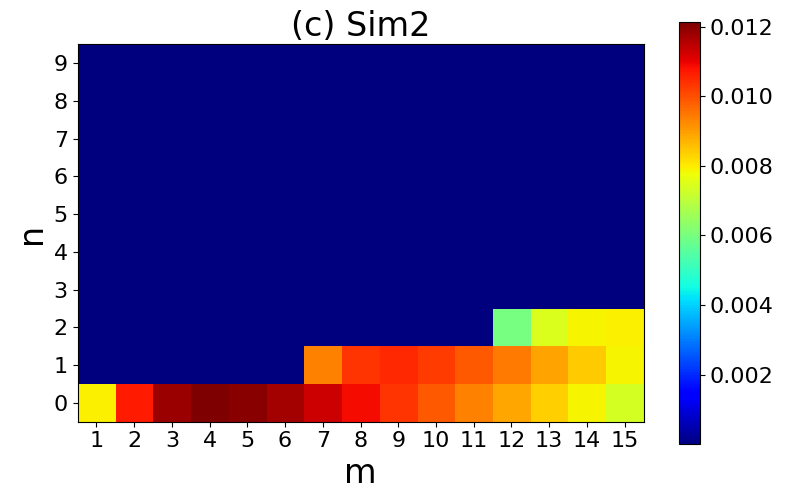}
   \includegraphics[width=0.3\textwidth]{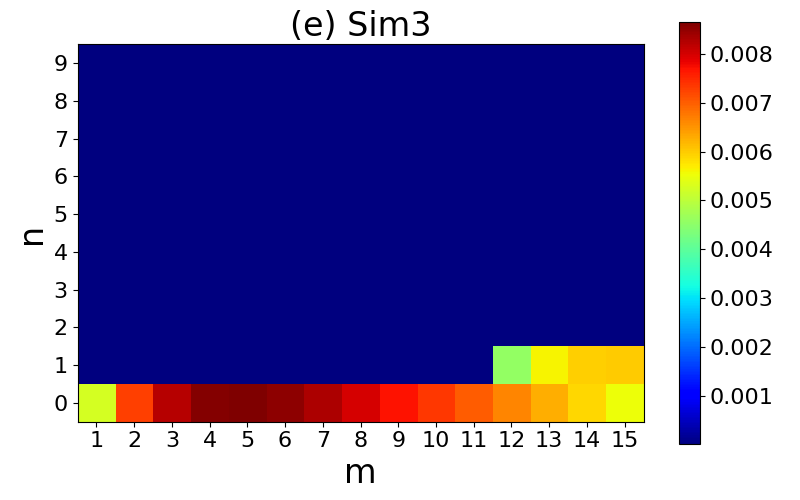}
    \caption{The growth rate ($s^{-1}$) for  modes of the initial equilibrium in Simulation SS-3D (Sim1, left), MS-3D (Sim2, middle), WS-3D (Sim3, right). Figure taken from \citet{2020ApJ...891...62L}.
    \label{fig:theory_rates}}
\end{figure}

The initial evolution is presented in more detail in \cite{2020ApJ...891...62L} but recapped here. The initial current sheet in each simulation is unstable to the tearing mode, but the varying reconnecting field strength and dimensionality results in slightly different behavior, based on the most unstable modes in the system \citep{Baalrud_2012}. Figure \ref{fig:theory_rates}, taken from \cite{2020ApJ...891...62L} shows the linear growth rate of the various modes in the initial 3D system for the choice of initial magnetic field and resistivity described above. The modes can be characterized by their wavenumbers (m,n) in the $x,z$ directions. As discussed in \citet{Baalrud_2012}, and in \citet{2020ApJ...891...62L}, the 
$n=0$ modes are parallel modes that are present in both 2.5D and 3D, occurring at the center of the current sheet ($y=0$), while the $n>0$ modes are oblique modes that occur in planes offset from the center ($y \ne 0$).

For strong reconnecting field, SS-3D, or Sim1 in \citet{2020ApJ...891...62L}, left panel, the most unstable mode is the (4,0) parallel mode, but there is also a strong $m=1$ oblique mode (5,1) with a similar growth rate. As the initial reconnecting field strength decreases (middle and right panels) all the modes become slower, but especially the oblique modes. Many low order oblique modes no longer exist. Thus, the system becomes more 2.5D like as the relative strength of the guide field, which remains fixed, becomes stronger as the reconnecting field decreases. Thus one expects that the weak reconnecting field 3D simulation (WS-3D) to be more 2.5D-like and hence similar to its 2.5D counterpart (WS-2D) than the strong reconnecting field simulation (SS-3D and SS-2D). 

\begin{figure}
    \centering
    \includegraphics[width=0.22\textwidth]{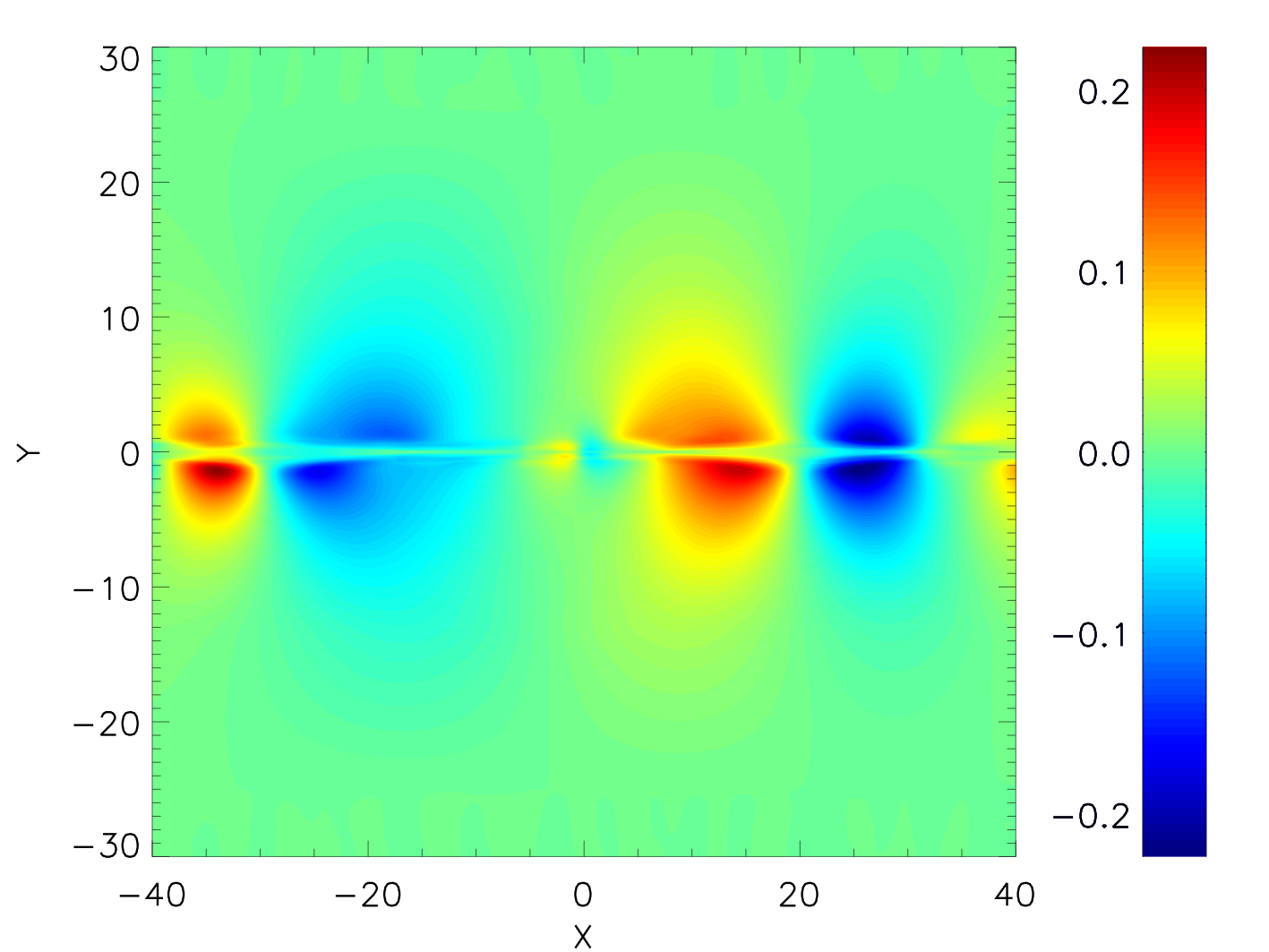} 
    \includegraphics[width=0.22\textwidth]{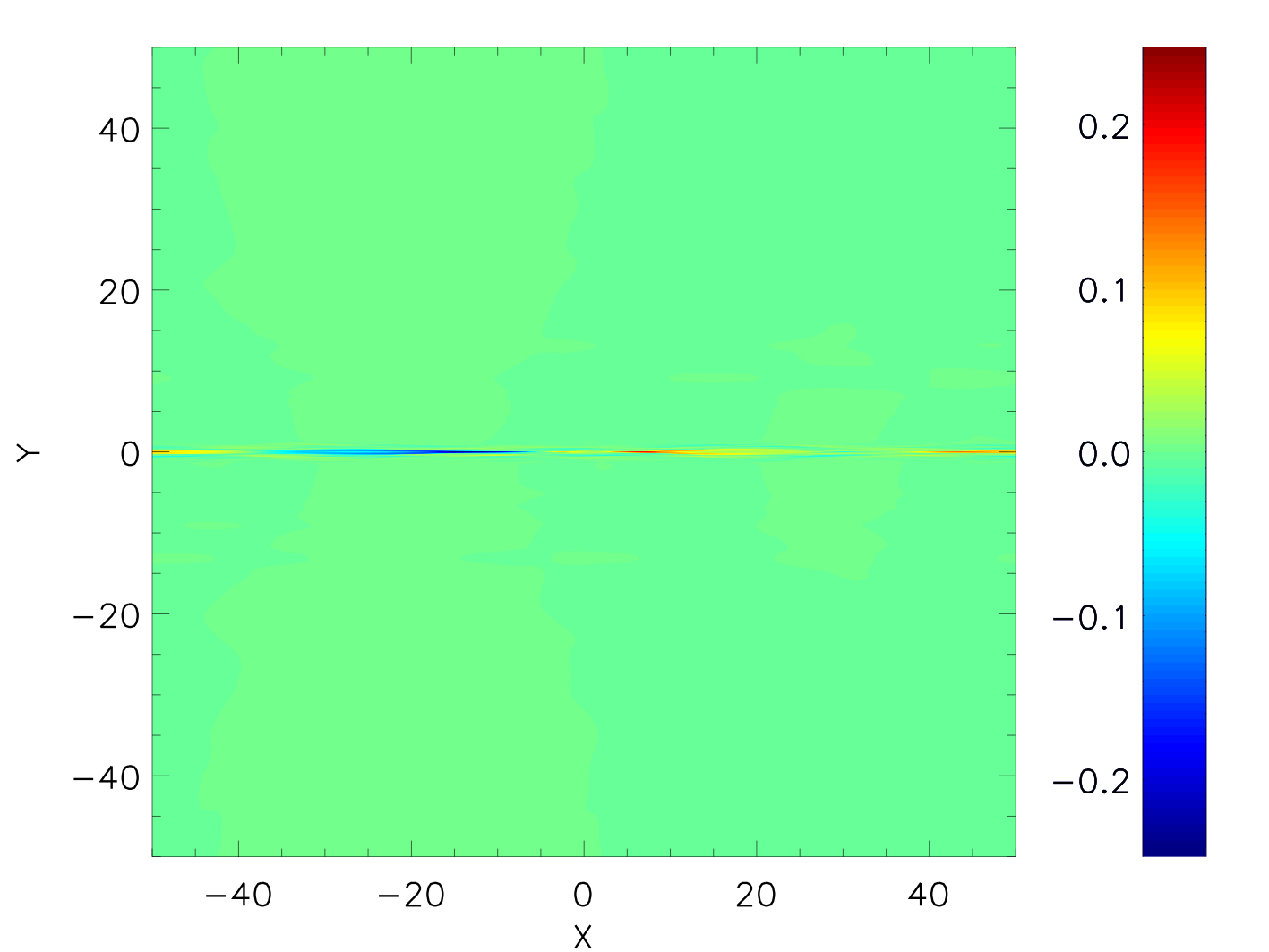} 
    \includegraphics[width=0.22\textwidth]{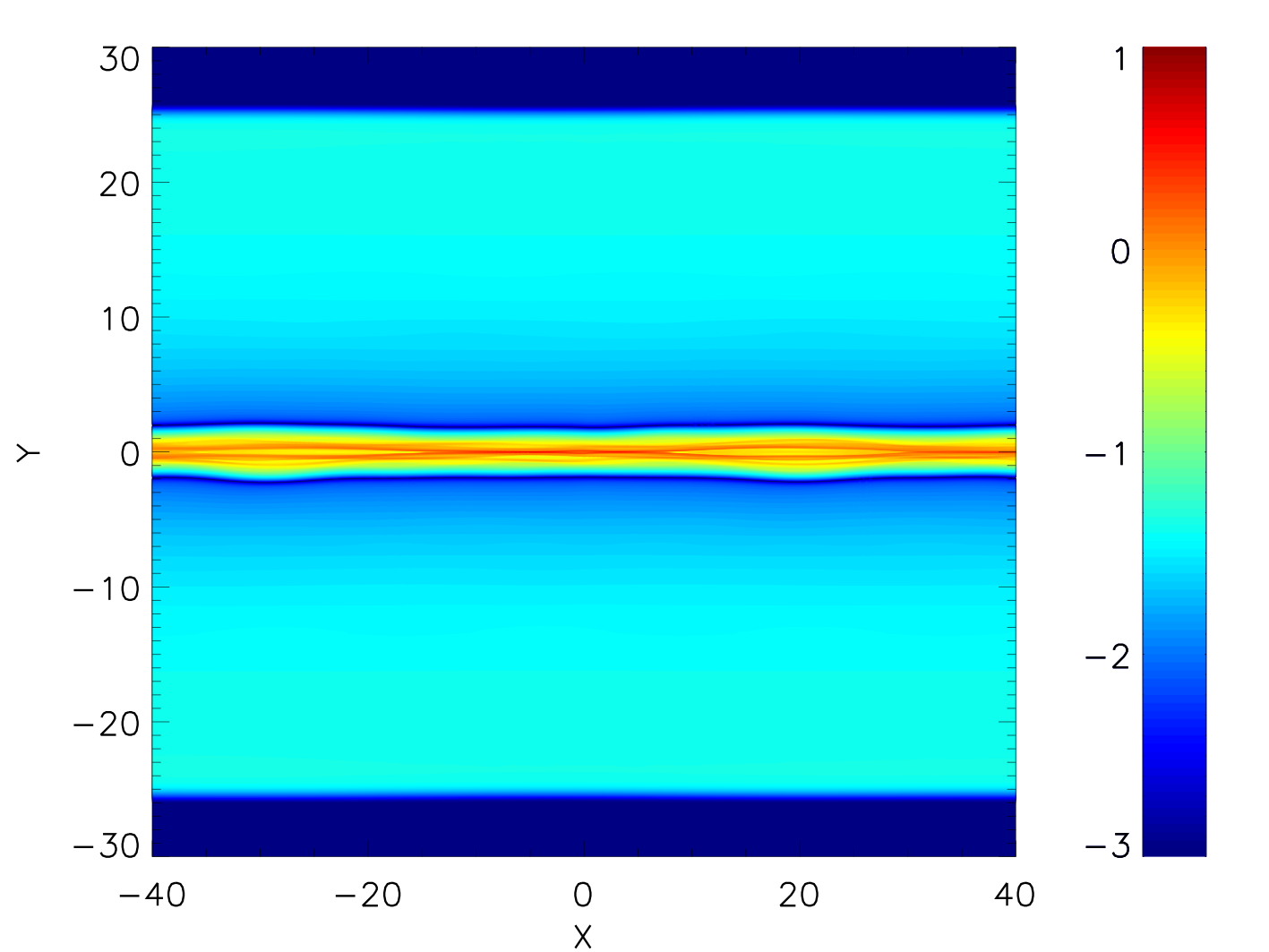} \\
    \includegraphics[width=0.22\textwidth]{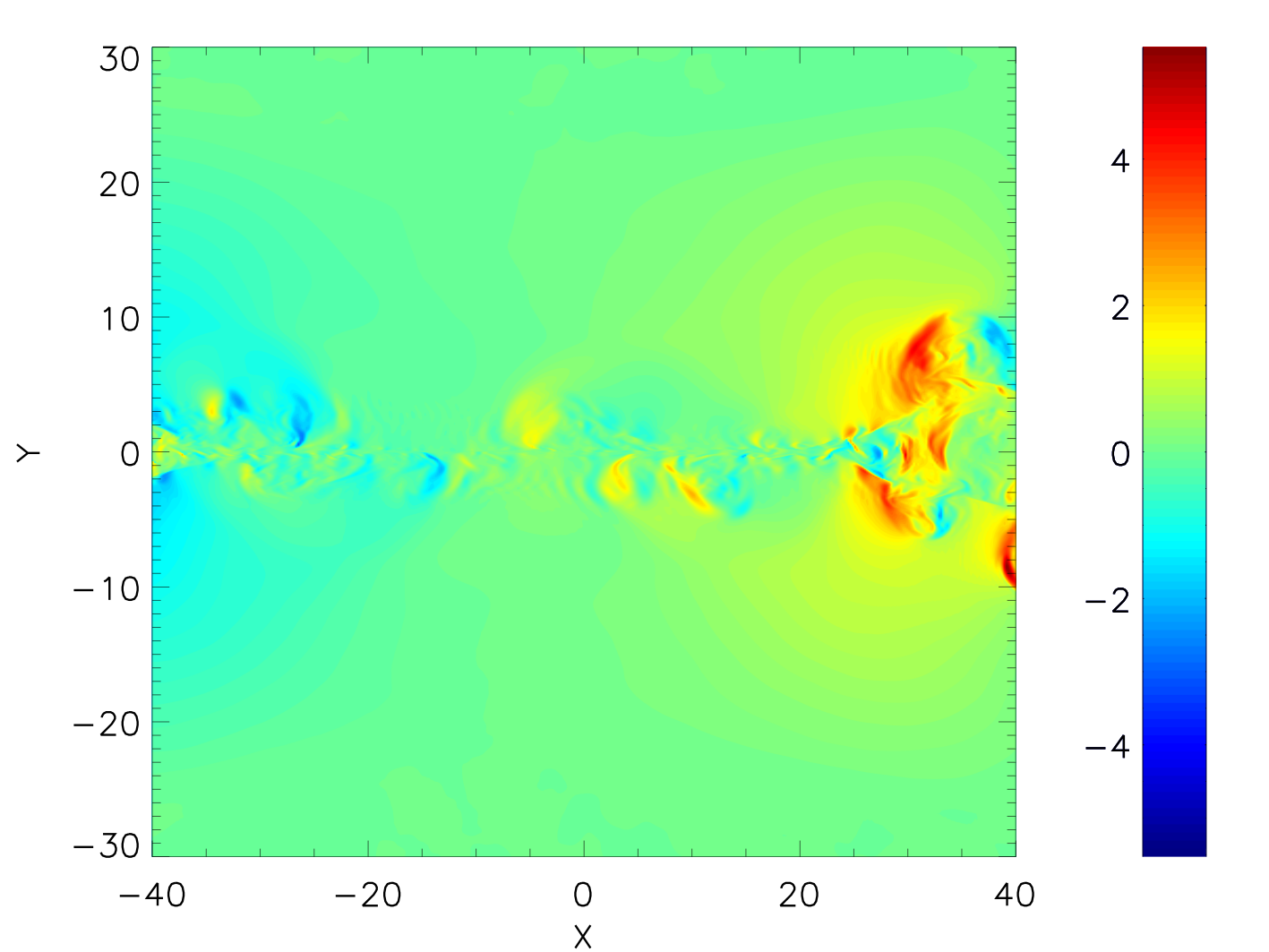} 
    \includegraphics[width=0.22\textwidth]{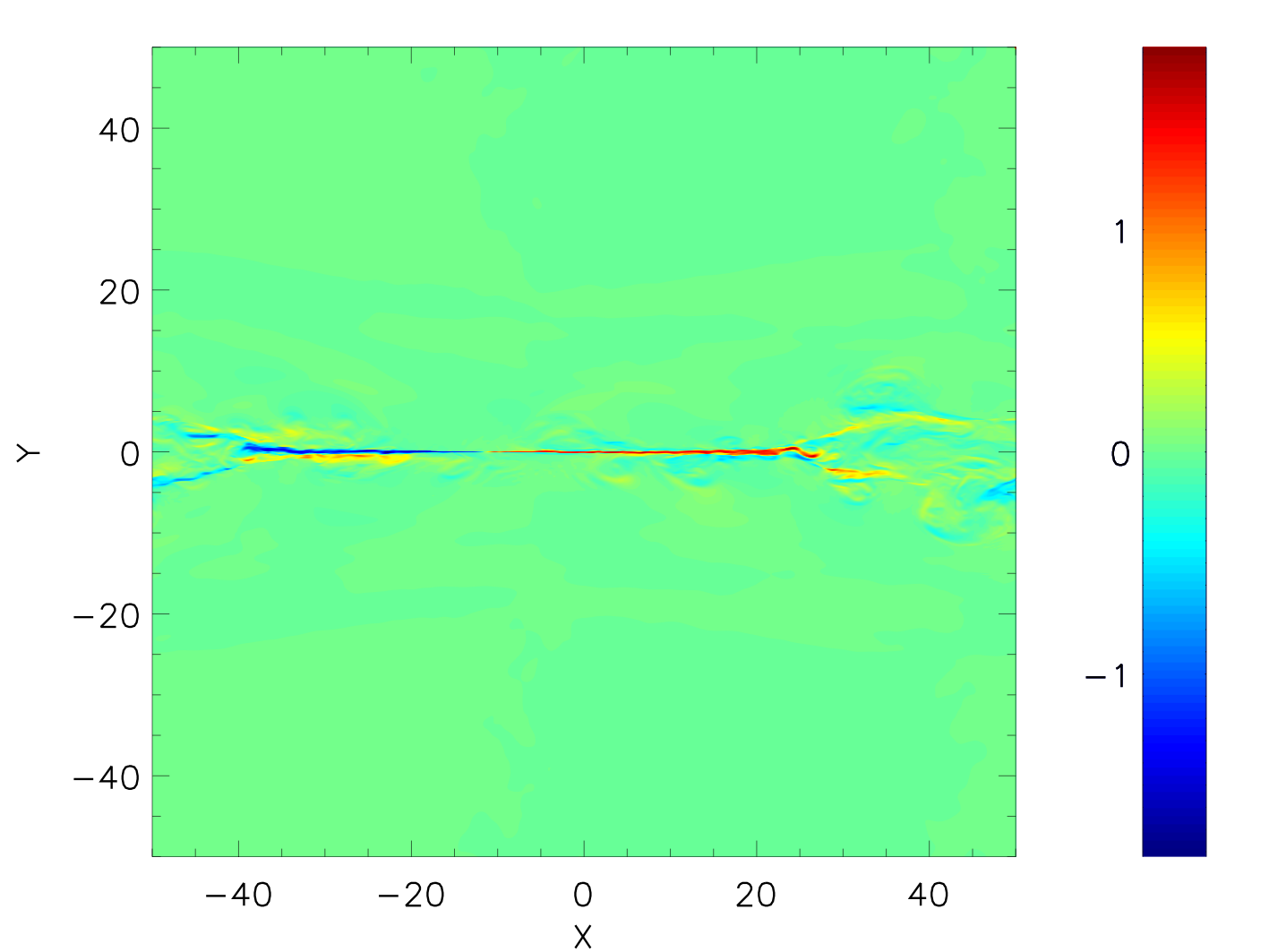} 
    \includegraphics[width=0.22\textwidth]{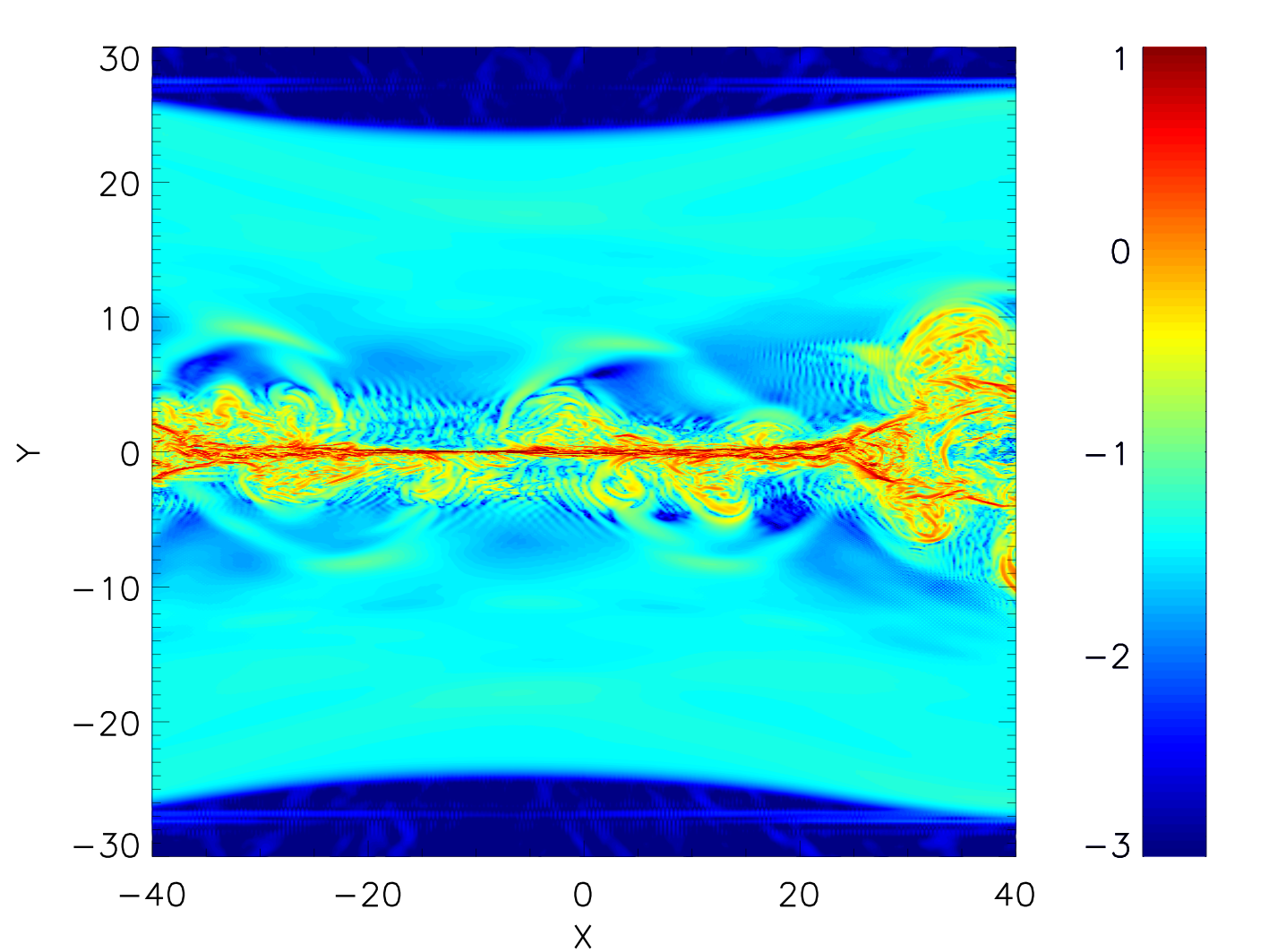} \\
    \includegraphics[width=0.22\textwidth]{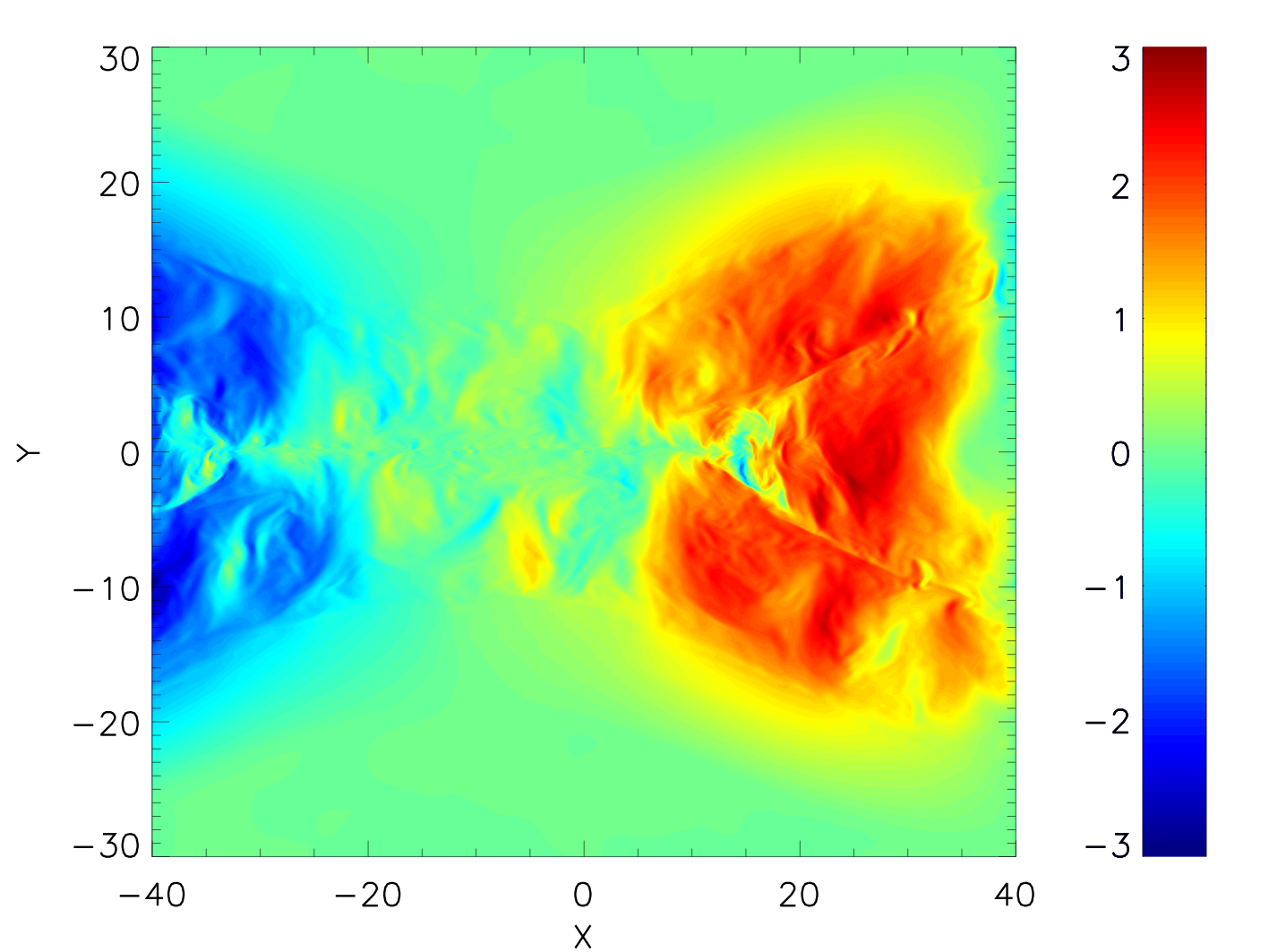} 
    \includegraphics[width=0.22\textwidth]{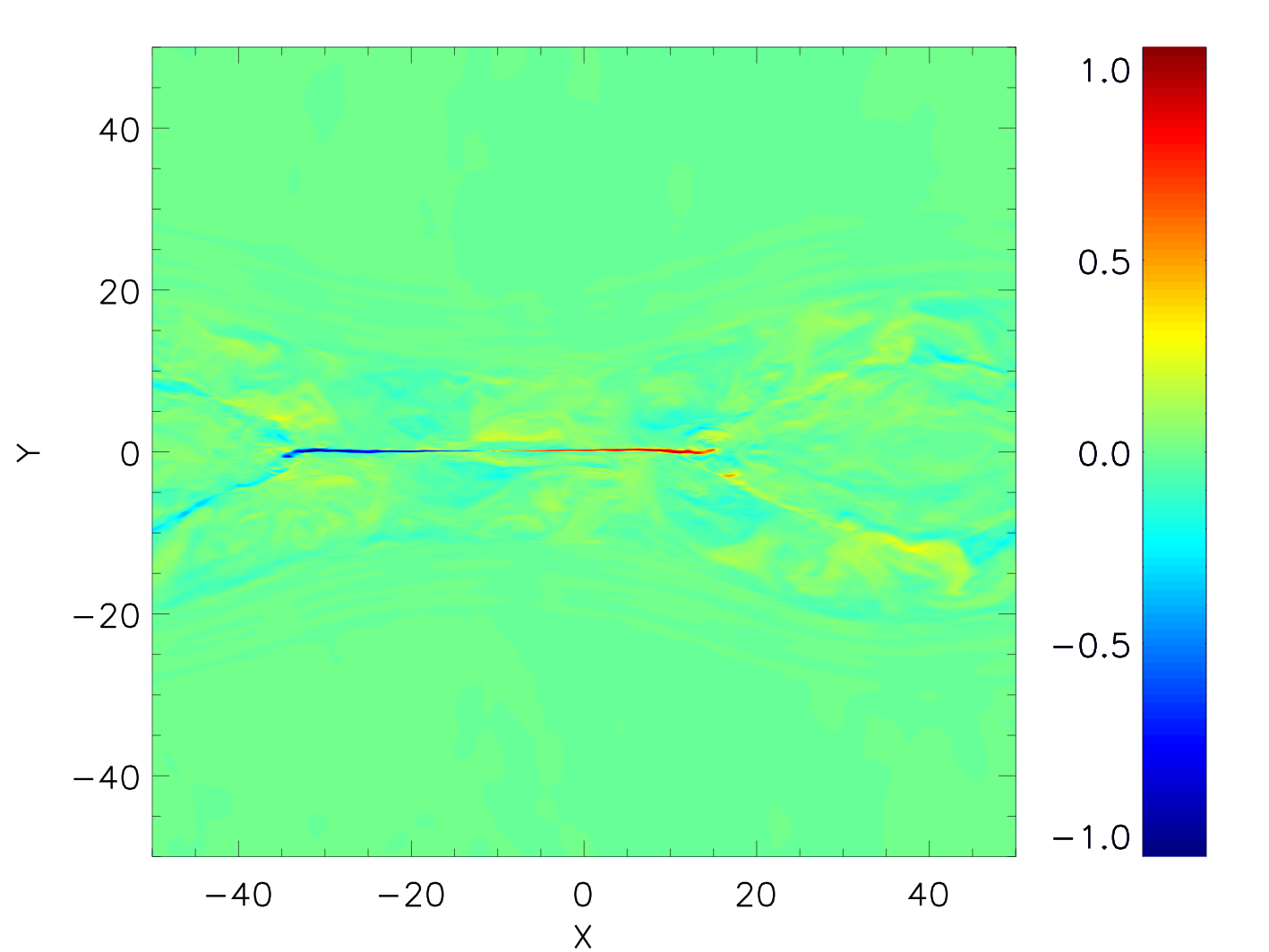} 
    \includegraphics[width=0.22\textwidth]{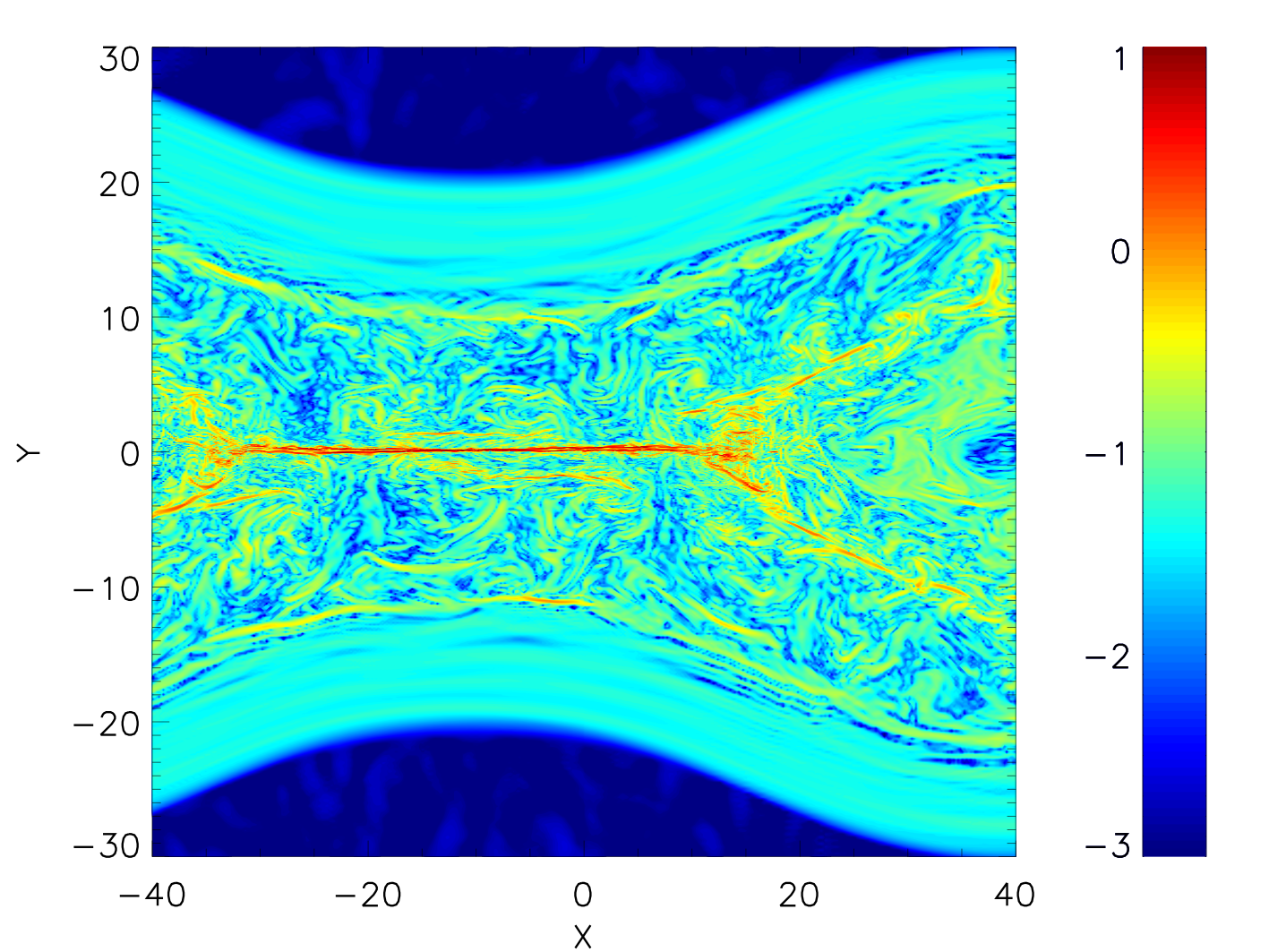}
    \caption{Initial evolution in the $z = 0$ plane of Simulation SS-3D at times [782,1000,1478] s, showing the coalescence of the initially dominant linear parallel and oblique modes to form a single flux rope and a primary current sheet connecting the two halves of this flux rope in the periodic system. Left panels: Reconnected field $B_{y}$ in G. Middle panels: Horizontal (outflow) velocity normalized by $V_0$. Right panels: $\log_{10}{|\mb{J}/J_0|}$. The $x$ and $y$ axes are in Mm. The data has been shifted in the periodic $x$ direction to place the new current sheet near the center of the plots.
    \label{fig:Sim1_3D_evolution}}
\end{figure}

\begin{figure}
    \centering
    \includegraphics[width=0.22\textwidth]{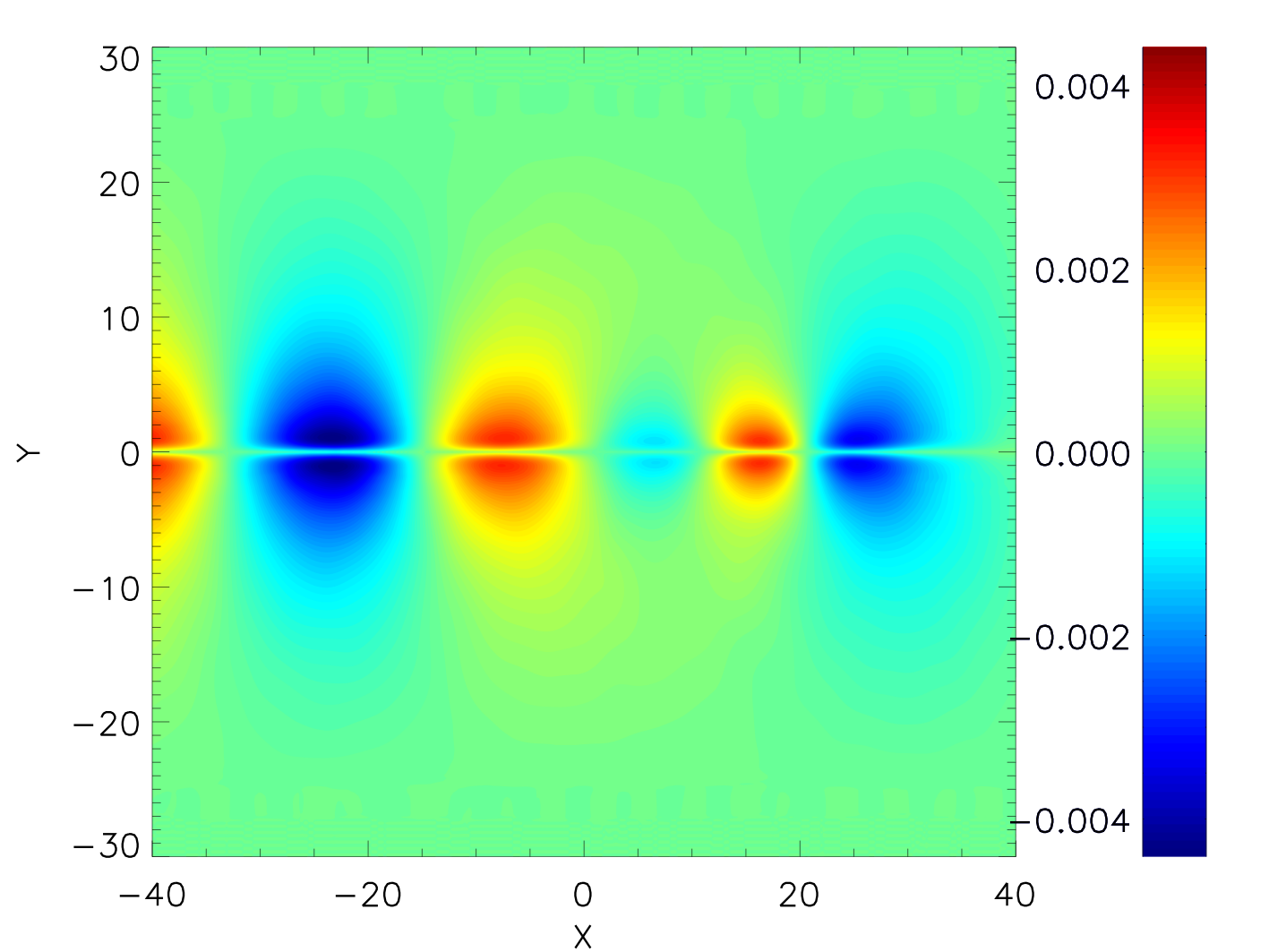} 
    \includegraphics[width=0.22\textwidth]{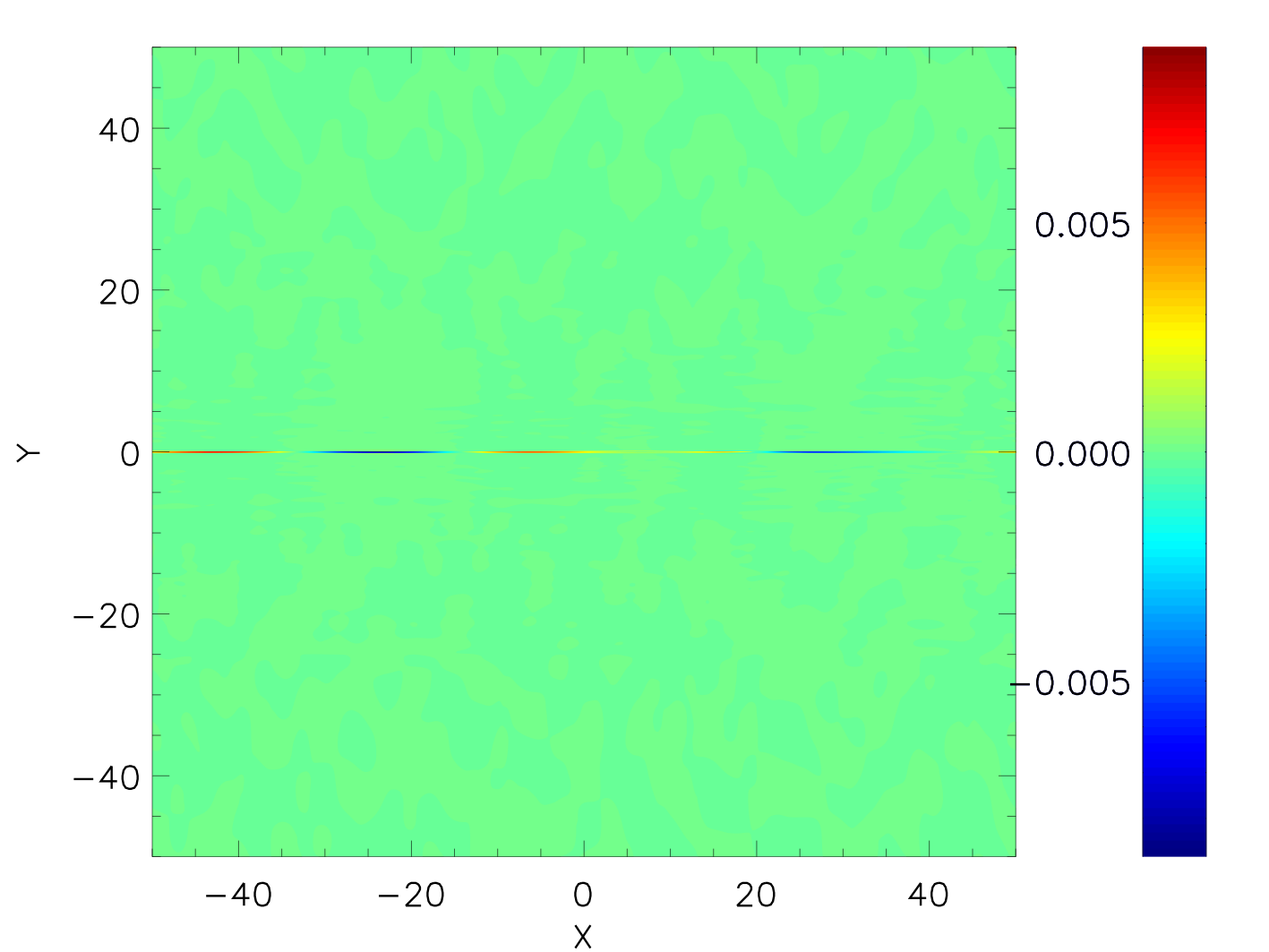} 
    \includegraphics[width=0.22\textwidth]{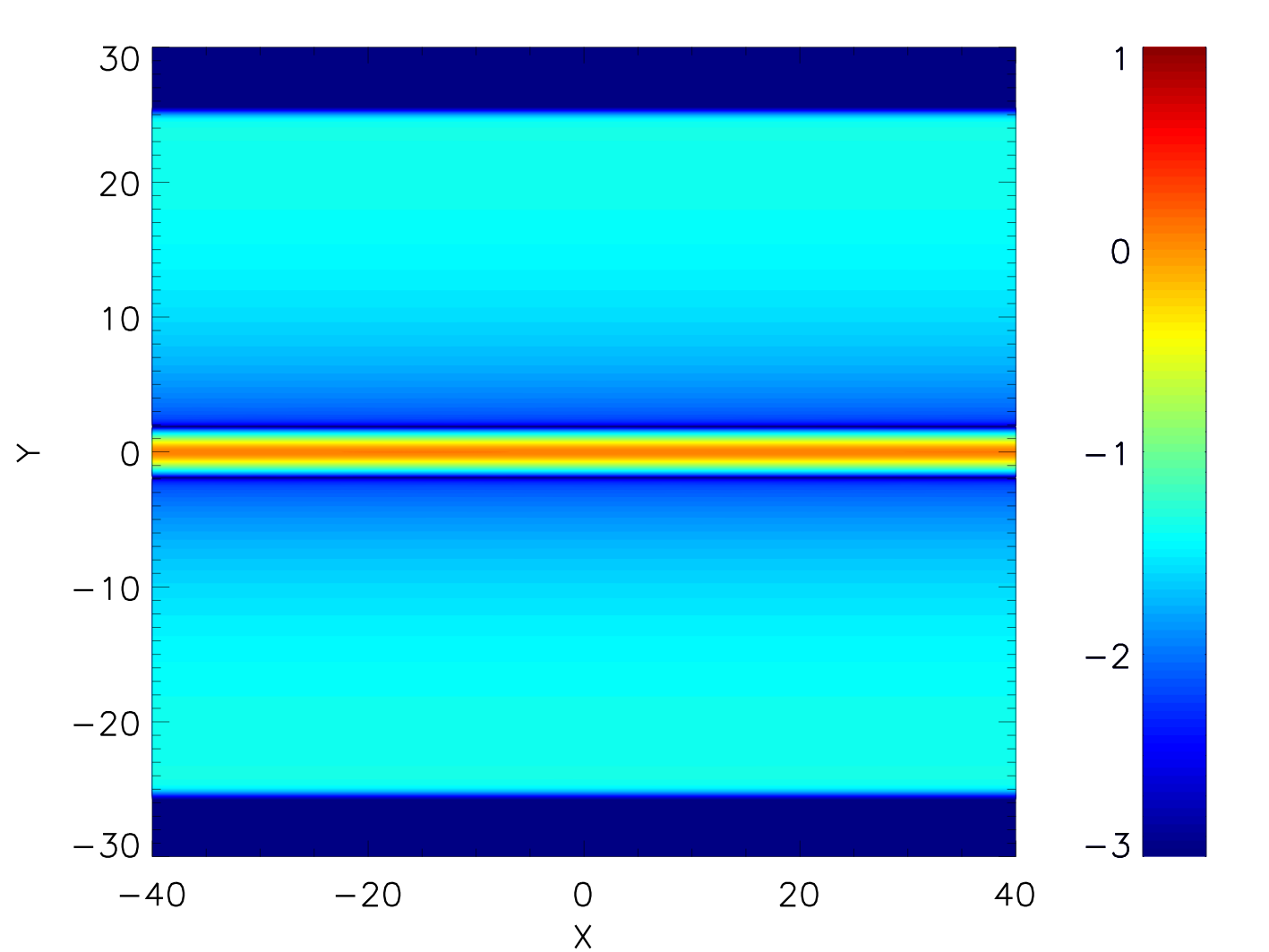} \\
    \includegraphics[width=0.22\textwidth]{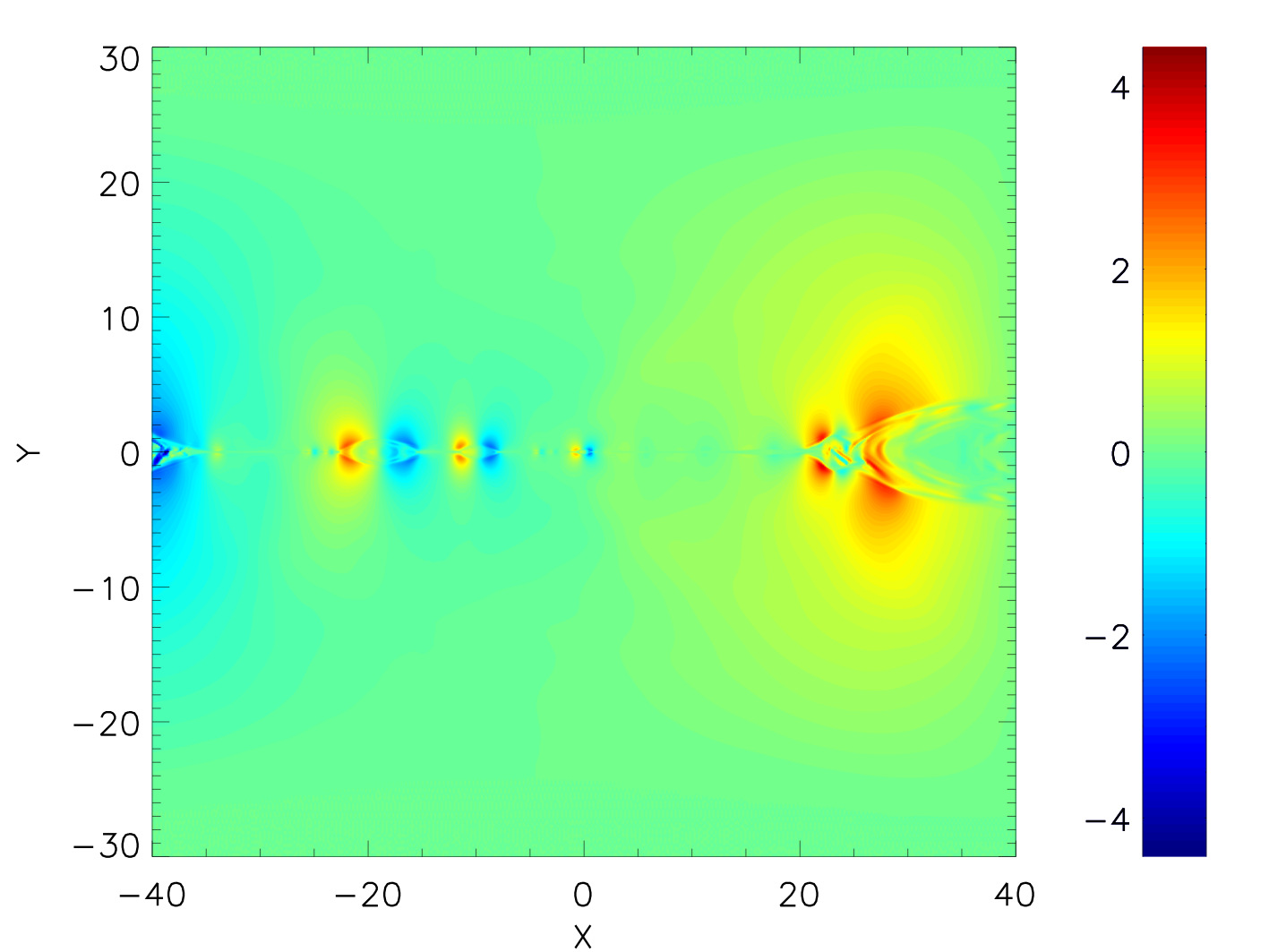} 
    \includegraphics[width=0.22\textwidth]{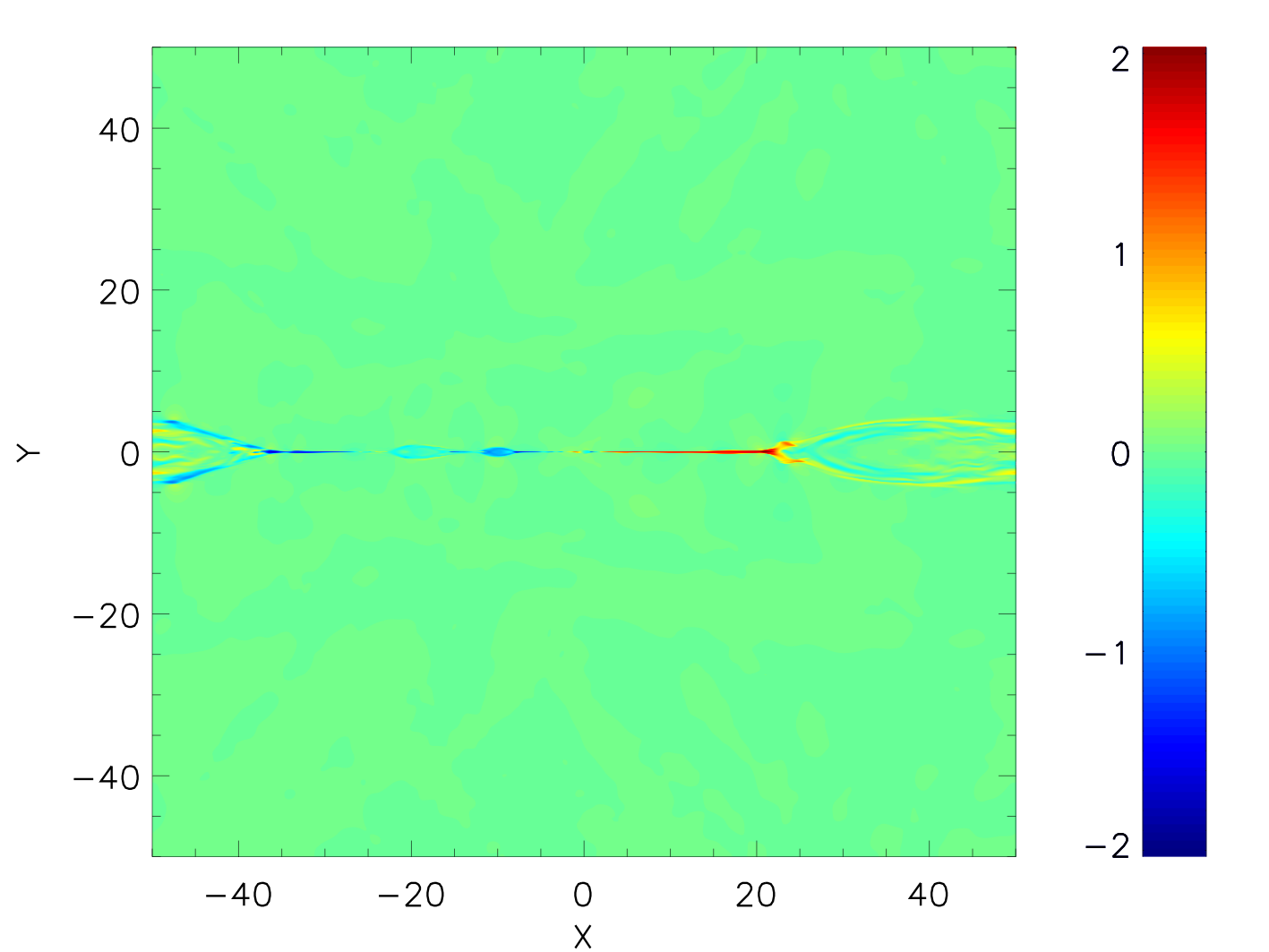} 
    \includegraphics[width=0.22\textwidth]{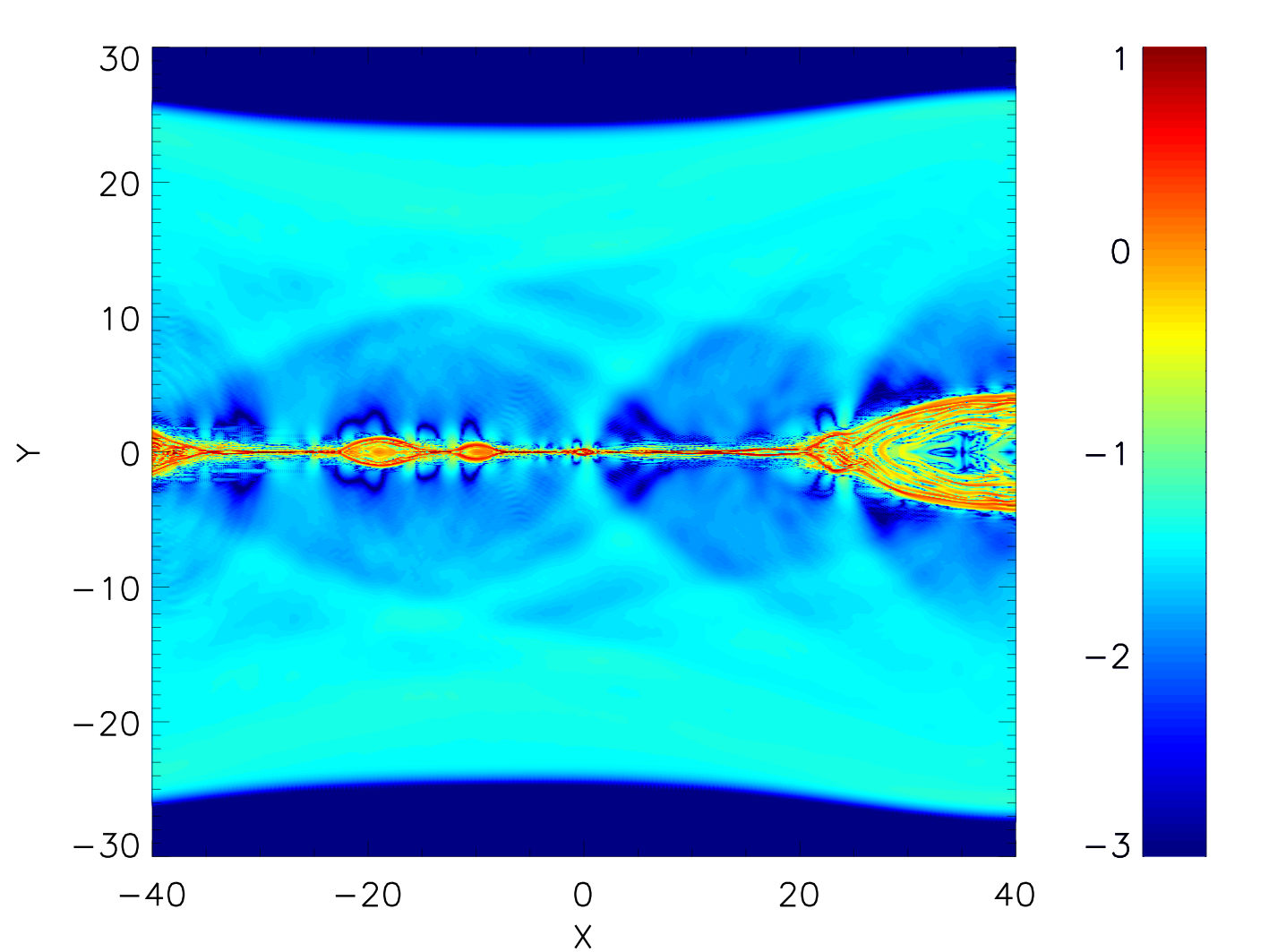} \\
    \includegraphics[width=0.22\textwidth]{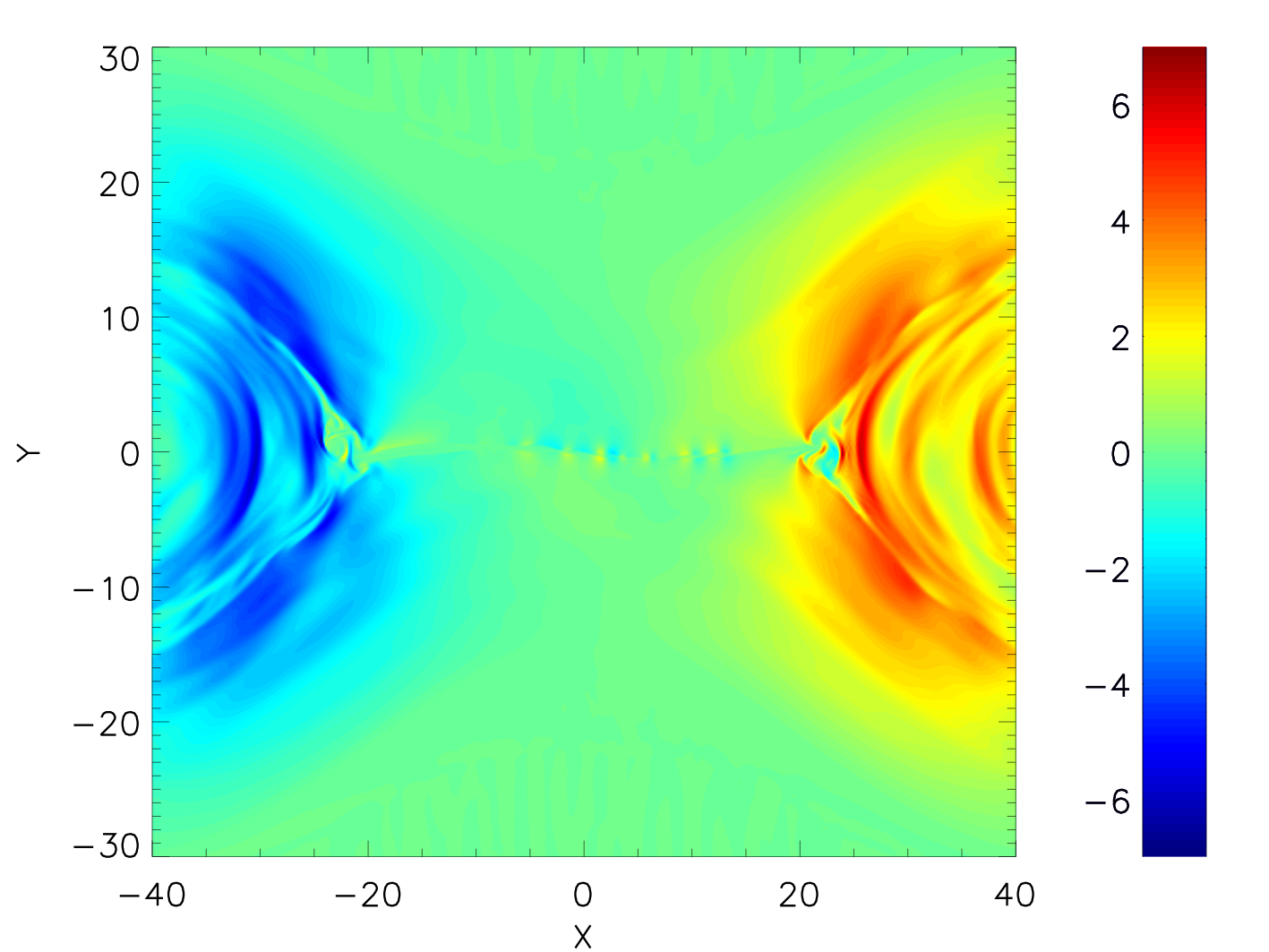} 
    \includegraphics[width=0.22\textwidth]{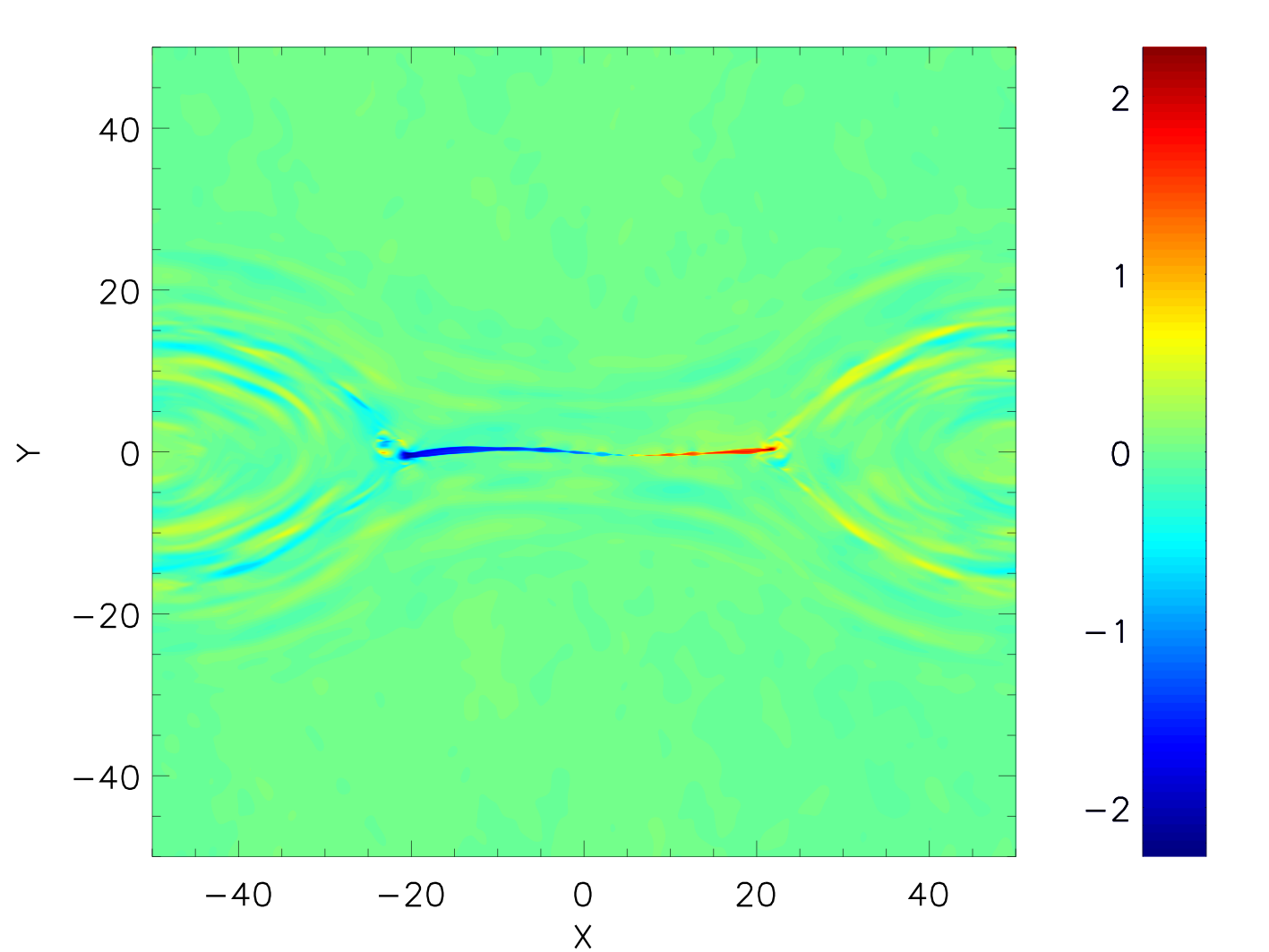} 
    \includegraphics[width=0.22\textwidth]{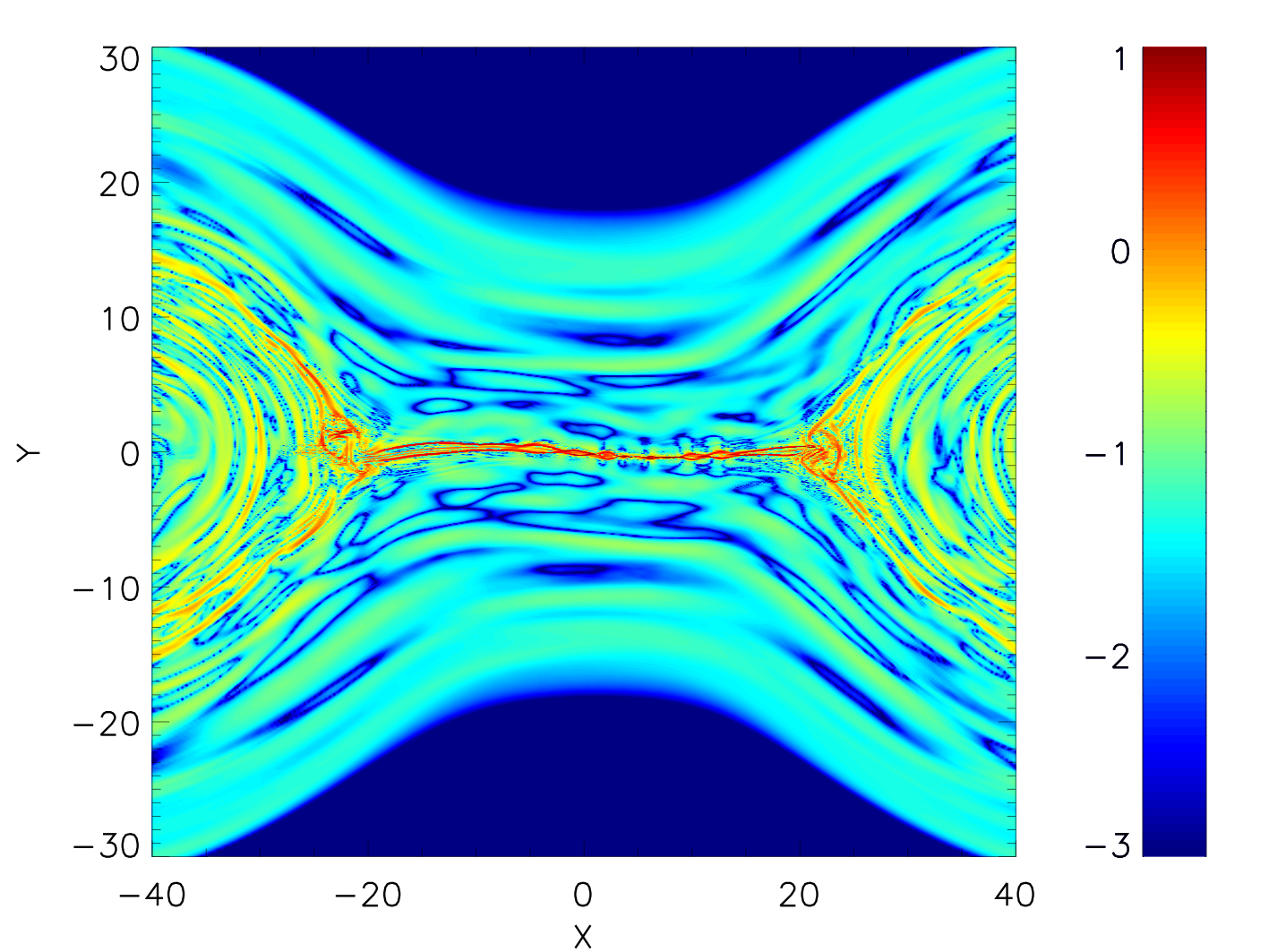} 
    \caption{Initial evolution of Simulation SS-2D. Same as Figure \ref{fig:Sim1_3D_evolution} but for the SS-2D at times [434,913,1130] s, and the right panels show $\log_{10}{|\mb{J_z}/J_0|}$.
    \label{fig:Sim1_2D_evolution}}
\end{figure}

In \cite{2020ApJ...891...62L}, the initial evolution of the simulations was shown to match well the expected superposition of the linear growth of the most unstable modes, which suggested the numerical model was matching the theoretical evolution well. Once the system became non-linear, comparison with linear growth was not  meaningful, but confidence was established in the accuracy of the numerical solution by the linear comparison.

In all simulations presented here, the general evolution is very similar. The unstable parallel modes, typically (4,0) dominate the early evolution,  as shown in Figure \ref{fig:Sim1_3D_evolution} for Simulation SS-3D (Sim1). However, the growth of parallel modes with lower $m$ (e.g., (2,0)) cause a coalescence of the flux ropes associated with the (4,0) mode into one large flux rope that is connected in the periodic system by a new primary current sheet. The reader is directed to the Figures in \citet{2020ApJ...891...62L} for a visual comparison of this process for Simulations SS-3D (Sim1), MS-3D (Sim2), and WS-3D (Sim3).

The presence of oblique modes in 3D, which manifest in reconnected field away from the (y=0, z=0) neutral line, add additional structure to the region just upstream from this newly formed primary current sheet. This can be seen in the $\log_{10}(J)$ plots (right panels) in Figure \ref{fig:Sim1_3D_evolution}, where the oblique modes can be seen, and which are not present in the 2.5D simulations, Figure \ref{fig:Sim1_2D_evolution}. It is the additional complex structure created by the oblique modes that affects later evolution of the newly formed primary current sheet. 

The 2.5D simulations evolve in a qualitatively similar manner, but as there is no variation along the ignorable direction $z$, there are no oblique modes present, as shown in Figure \ref{fig:Sim1_2D_evolution}. As a result the coalescence process which forms the large flux rope and the new primary current sheet is unencumbered by the interacting oblique modes, and the system in general is much more laminar and less complex. 

\begin{figure}
    \centering
    \includegraphics[width=0.5\textwidth]{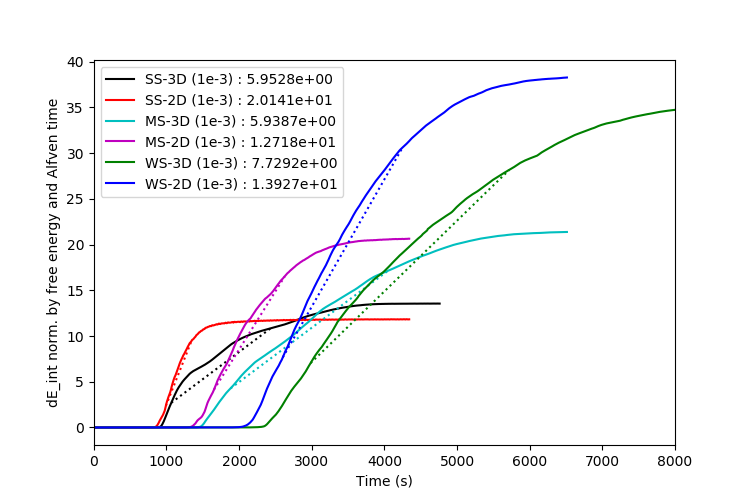}
    \caption{Change in internal energy, normalized by $E_{free}/t_{A}$,  for all 6 simulations.
    \label{fig:internal_energy}}
\end{figure}

The change in internal energy is shown in Figure \ref{fig:internal_energy} for all 6 simulations. The internal energy has been normalized by $E_{free}/t_{a,0}$ where $E_{free}$ is the available free magnetic energy at the start of the simulation, and $t_ {a,0}=L_{x}/V_{a,0}$  where the Alfv\'{e}n speed is based on the reconnecting field component, $B_{x,0}$. This normalization will create a dimensionless reconnection rate when the time derivative of this quantity is taken (see later in \S \ref{sec:results_rates}). For a given reconnecting field value, it is clear that the heating rate is faster for 2.5D compared to 3D, as can be seen by the gradients in these plots, and the values of the gradient shown (slope of dotted line). The reason for this will be discussed in \S \ref{sec:results_turbulence}.

The main temporal period of internal energy increase in these simulations starts at the later stages of the linear mode coalescence and ends well after the new primary current sheet is formed. In later stages, the majority of the heating is caused by the breakup of the newly formed primary current sheet, as it is unstable to further tearing, as in the cascading plasmoid evolution presented in \citet{Huang_2013}. However, this tearing instability  has a pre-condition dependent on the earlier evolution and affected by oblique modes for the 3D simulations. 

 The remaining analysis focuses on the large scale evolution of this newly formed reconnecting, tearing current sheet, rather than looking at individual reconnection sites, of which there are many, due to the interaction of multiple tearing modes in 3D.

%% file: results_turbulence.tex
\subsection{Oblique Mode Generated Structure: Effect on the Reconnection Rate}
\label{sec:results_turbulence}

\begin{figure}
    \centering
    \includegraphics[width=0.45\textwidth]{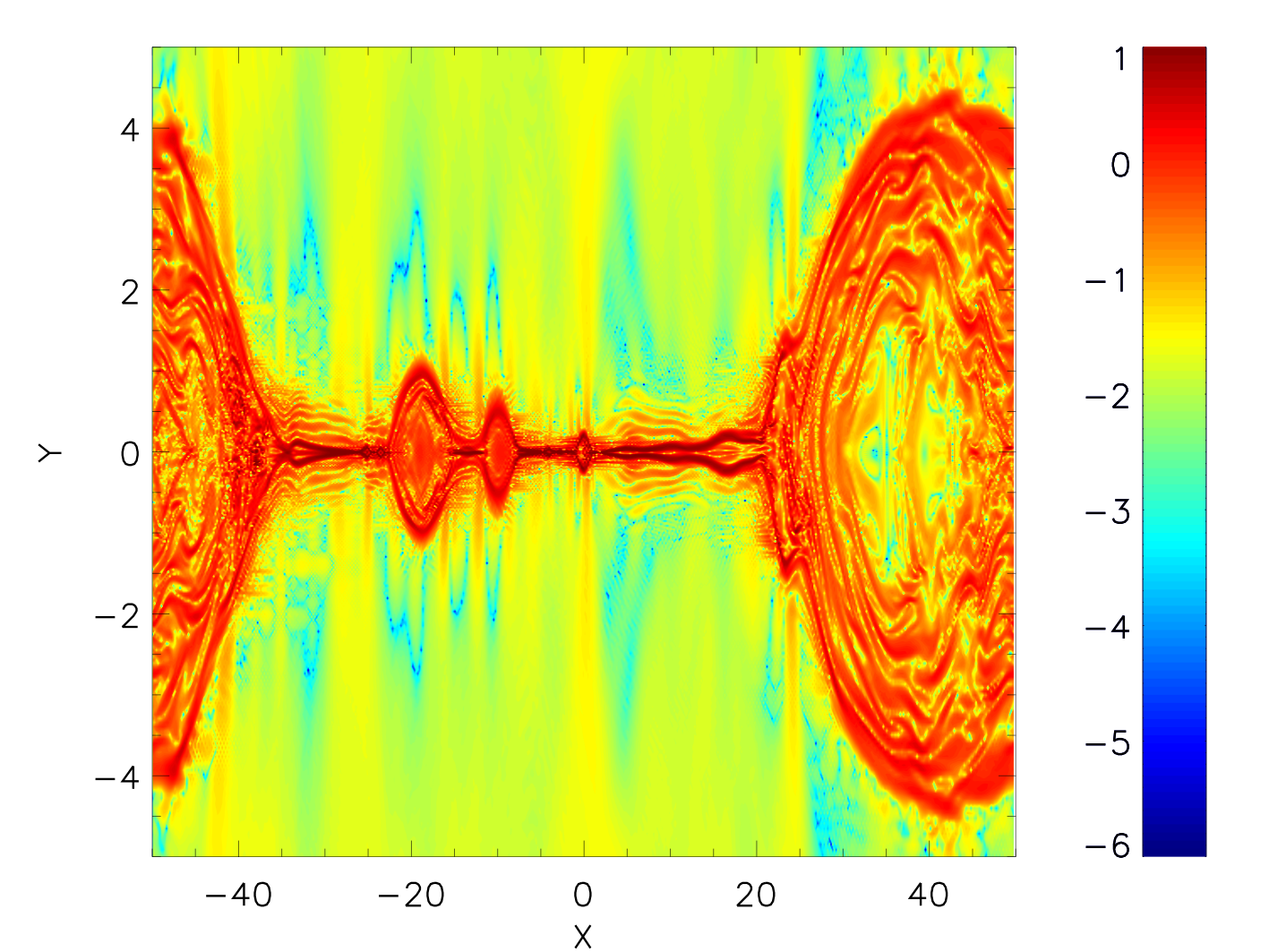}
     \includegraphics[width=0.45\textwidth]{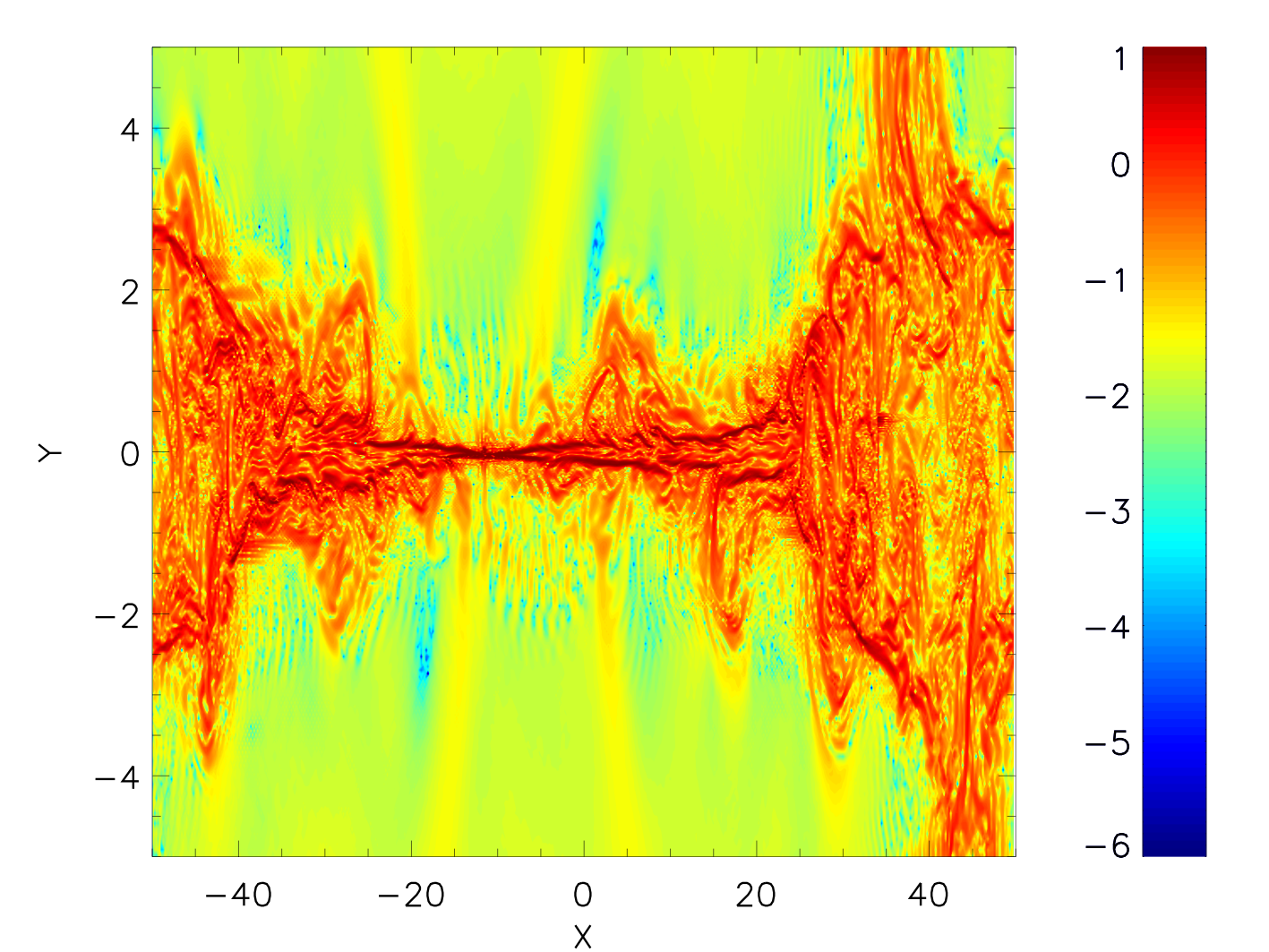}
    \caption{$\log_{10}{|J_{z}|}$ in Simulation SS-2D (left) at time 913 s, and Simulation SS-3D (right) at time 956 s. Plasmoid evolution is much more coherent in the 2.5D simulation (left).
    \label{fig:2D_vs_3D_plasmoids}}
\end{figure}

In \S \ref{sec:results_initial}, Figures \ref{fig:Sim1_3D_evolution} and \ref{fig:Sim1_2D_evolution} it was shown that the presence of oblique modes in 3D adds more complex structure to the newly formed current sheet, compared to the 2.5D system. This complexity affects the tearing of the newly formed current sheet. Figure \ref{fig:2D_vs_3D_plasmoids} shows a snapshot of $\log_{10}{|J_{z}|}$ in the 2.5D and 3D simulation for the strong recconnecting field case. Note that the $y$-range is much smaller than in Figures \ref{fig:Sim1_3D_evolution} and \ref{fig:Sim1_2D_evolution}. The oft-observed presense of plasmoids being ejected to the ends of the current sheet \citep[e.g.,][]{Bhattacharjee_2009} are clearly seen in the 2.5D case (left panel). 

In the 3D case (right panel of Figure \ref{fig:2D_vs_3D_plasmoids}), the newly formed current sheet is much less coherent, and the effect of oblique modes away from $y=0$ can clearly be seen. Plasmoids in the 3D current sheet are less coherent and less numerous. Given that the ejection of these plasmoids and their interaction with the main flux rope at either end is the dominant source of energy dissipation (via viscous damping), this is why the rate of change of internal energy, i.e., the heating rate, is faster in the 2.5D simulations compared to the 3D simulations (\S \ref{sec:results_initial}, Figure \ref{fig:internal_energy}).

In 3D, non-linear oblique mode interaction generates a wide region of complex flows and currents, as seen in Figure \ref{fig:Sim1_3D_evolution}, bottom right panels. It is worth noting that \citet{Huang_2016} report on self-sustained turbulence created by oblique modes, and examined its effect on the plasmoid evolution in the current sheet. Here, the analysis of the complex inflow region created by interacting oblique modes with respect to turbulence is left for future publication, while the large scale evolution and overall reconnection rate is considered here.

The oblique mode effect on the current sheet evolution can be visualized by considering the spatial variation in $x$ of the current sheet parameters. At each point in $x$ and $z$, the reconnecting field profile $B_x(x,y,z)$ in the new current sheet is fitted to a function in $y$ with the same form as the original sheet.

\begin{equation}
    B_{x}(y)  =  {\mathit B_{x,0}}\tanh{\left(\frac{y - {\mathit c}}{\mathit a}\right)}\cos{(0.5 \pi \frac{y}{\mathit b})}, ~ \textrm{for} ~|y| < {\mathit b} \label{eqn:fittingBx}
\end{equation}

\begin{figure}
    \center
    a)
    \includegraphics[width=0.3\textwidth]{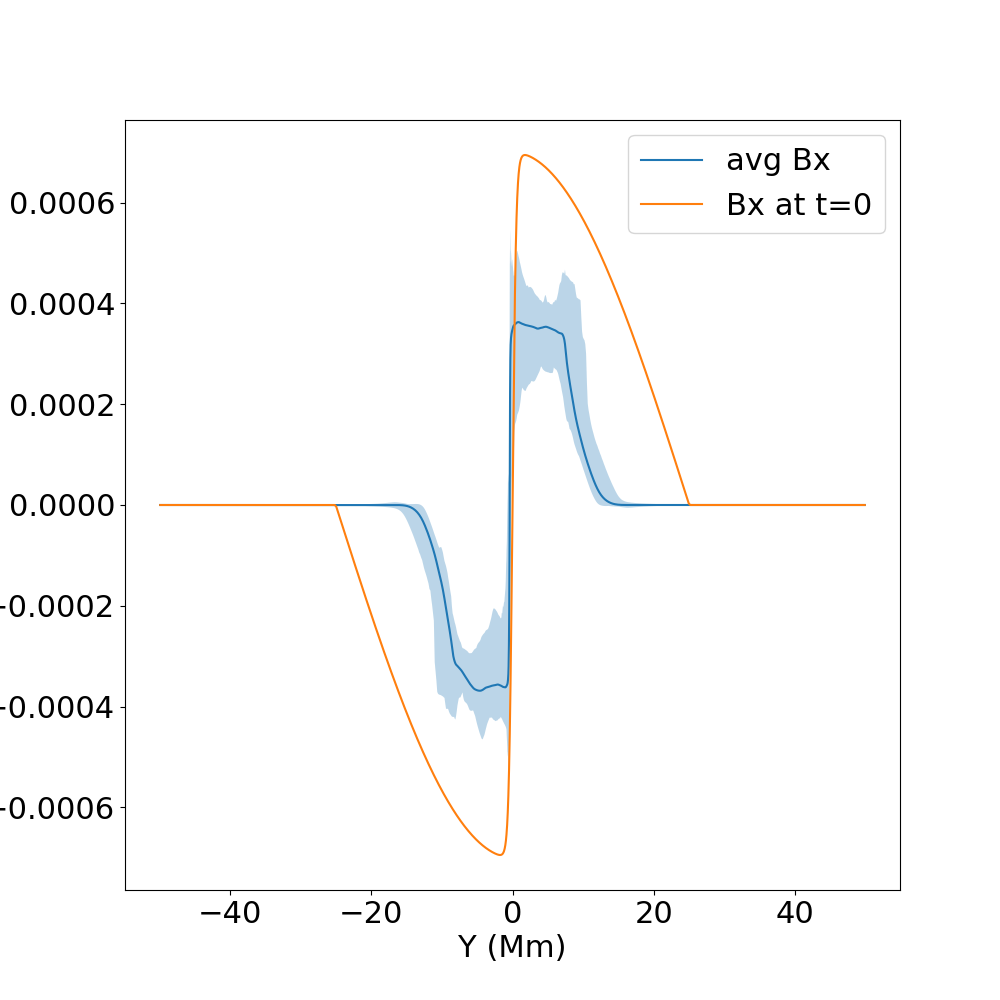}
    b)
    \includegraphics[width=0.3\textwidth]{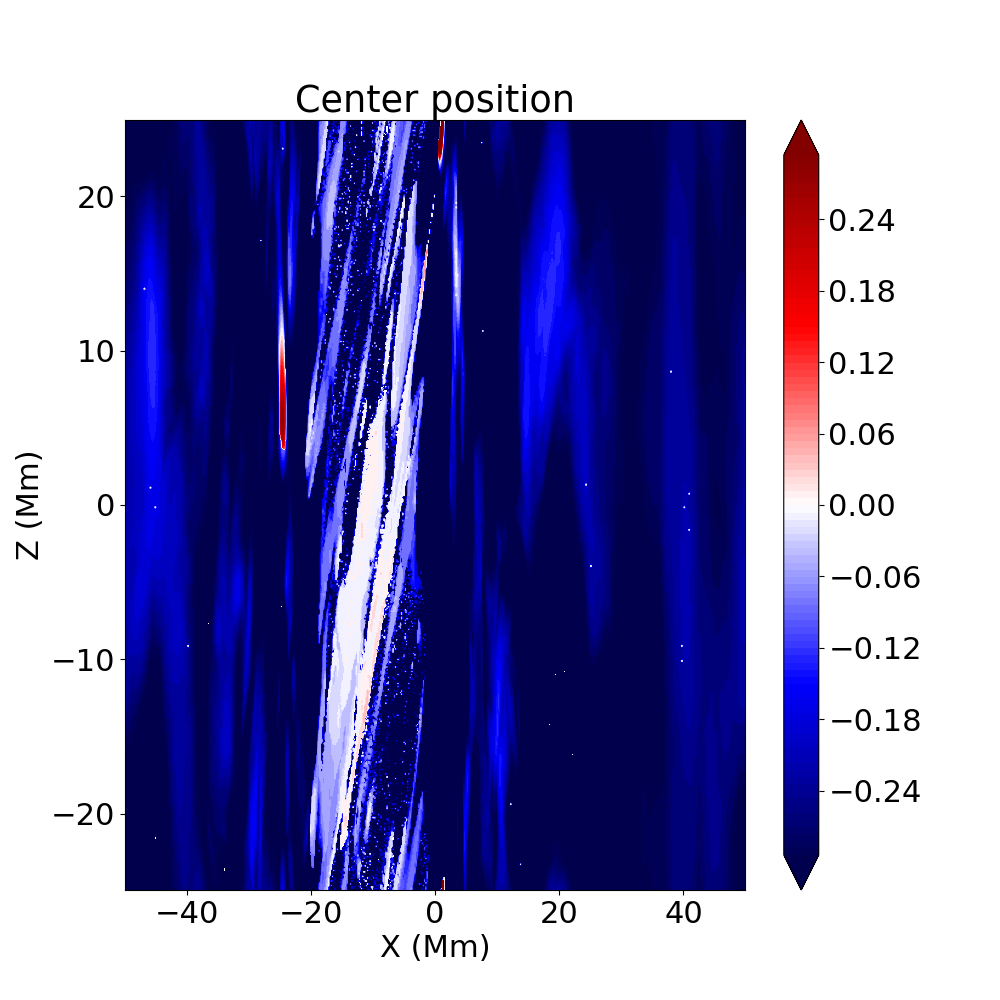}
    c)
    \includegraphics[width=0.3\textwidth]{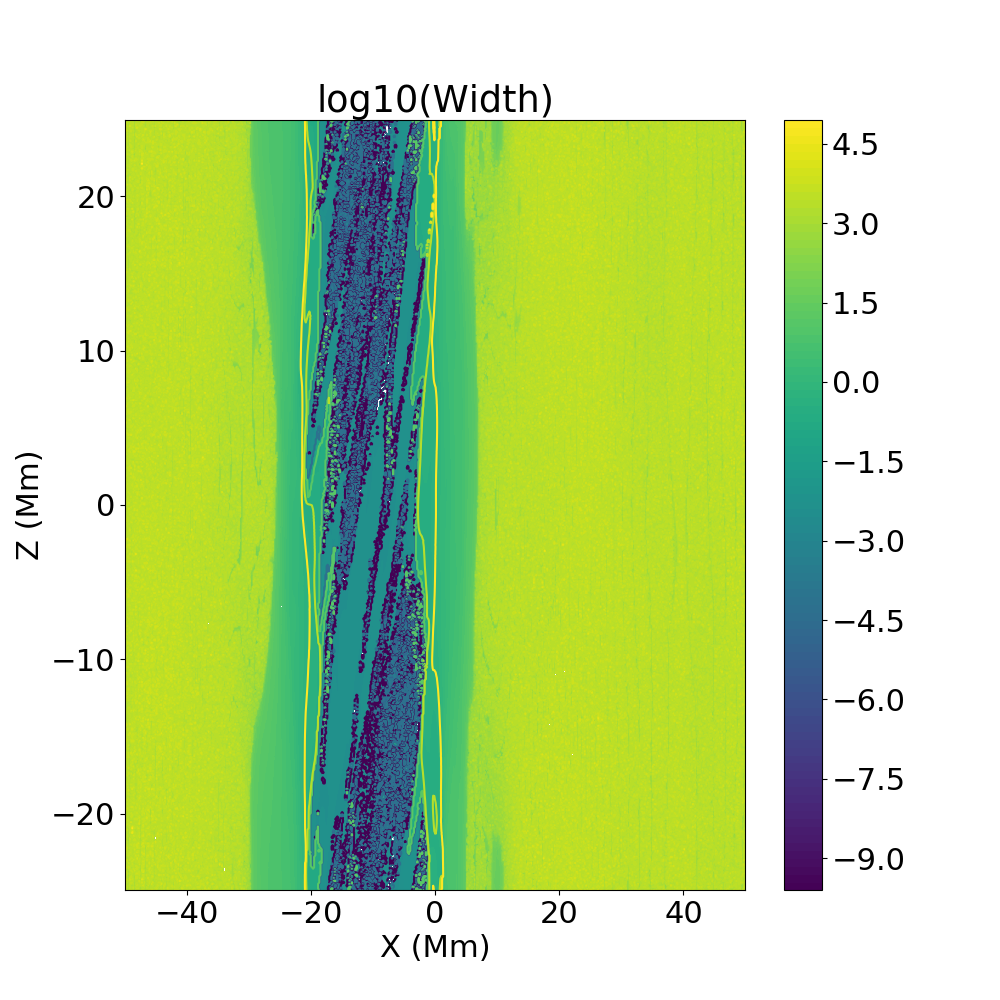}
\caption{Left panel: Orange line shows the initial $B_x(y)$ profile. The blue solid line shows the average $B_x(y)$ profile in the $x$ domain [-20,0] at time 2605 s for the strong reconnecting field simulations (SS-3D). The shaded blue region is the range of $B_x(y)$ in the same $x$ domain. Middle panel: The fitted parameter $c$ (position in $y$) as a function of $x$ and $z$. Right panel: The fitted parameter $a$ (width) as a function of $x$ and $z$. The center position and width are found by fitting the absolute value of $B_x$ to the absolute of equation \ref{eqn:fittingBx}. As in the earlier plots for SS-3D, the arrays in the middle and right plot have been shifted half a domain in $x$ to place the current sheet near the center of the plot.
\label{fig:NewCS}}
\end{figure}

Figure \ref{fig:NewCS}(a) shows the initial current sheet profile (orange line), and then the results of the fit in the $x$ domain where the current sheet is located, $x=[-20,0]$ Mm. The $x$-averaged profile in this sub-domain is the solid blue line, and the range over this $x$ sub-domain is the blue shaded region. Panels (b) and (c) of Figure \ref{fig:NewCS} shows the fitted center position $c$ and width $a$ in Eqn. (\ref{eqn:fittingBx}), as a function of $x$ and $z$. In these plots, vertical structures would manifest from mainly parallel tearing modes, while tilted structures would manifest from oblique modes. In the $x$ sub-domain outside of $x=[-20,0]$ Mm, where the large flux rope is located, the actual values of the fits are not relevant but the structure of the values, which are mainly vertical, reflect the parallel, 2.5D-like, nature of the evolution. In the $x$ sub-domain of $x=[-20,0]$ Mm where the primary current sheet is located, tilted structures are seen in both fitted parameters, indicating the effect of oblique modes. 

Based on the fact that interacting oblique modes in the 3D simulations appear to be the source of the complex structure in the newly formed current sheet, the  following interpretation is offered. Tearing modes are associated with flux tubes, the cross sections of which are identified as tearing islands or plasmoids. The tubes are aligned with the local magnetic field direction and therefore have different angles of obliquity as a function of distance from the center of the sheet. Reconnection in the 3D current sheet is therefore patchy, in both the reconnecting plane and the guide field direction. Consequently, tubes that have reconnected with neighbors are now interlinked. This restricts their ability to flow freely past one another. The otherwise organized behavior seen in 2.5D – formation of islands and ejection to the ends of the sheet - is complicated by this entangling of flux tubes, which impedes the inflow and outflow, thus affecting the plasmoid evolution, and as will be seen, the overall reconnection rate. This negative effect on the reconnection process is greater for large reconnecting field strength, because more oblique modes are present in the system, as shown in Figure \ref{fig:theory_rates}. For weak reconnecting field, the system is more 2.5D-like as there are much less oblique modes, and the effective guide field is strong. Note that this interlinking of flux tubes during reconnection has been suggested as a potential cause of reduced energy release during large solar eruptions \citep{Klimchuk_1996}. It also has implications for the magnetosphere \citep{Otto_1995,2019GeoRL..46.1937O}.

\begin{figure}
    \centering
    \includegraphics[width=0.33\textwidth]{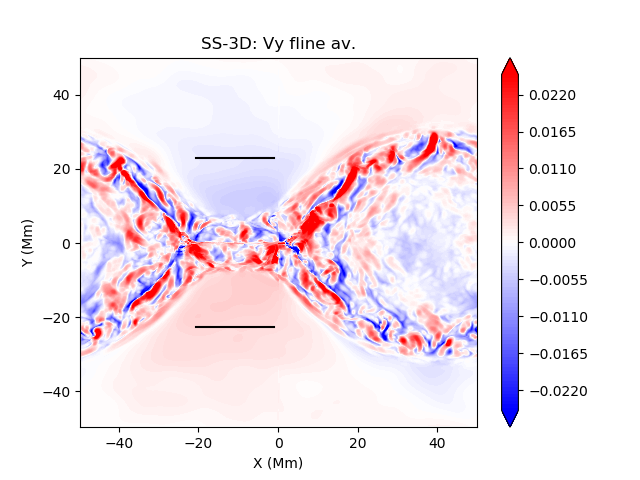}
    \includegraphics[width=0.33\textwidth]{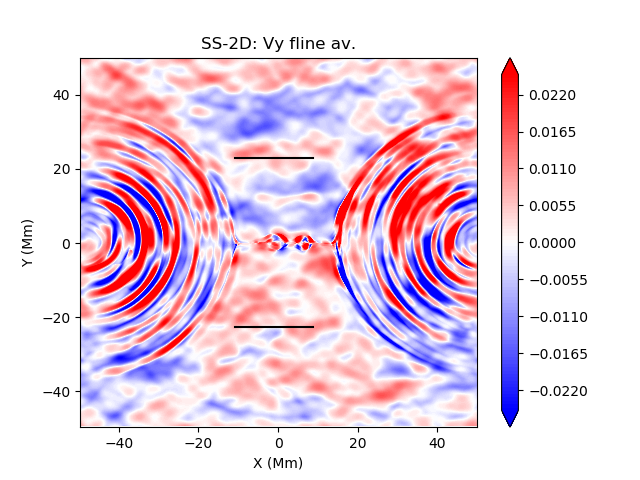} \\
    \includegraphics[width=0.33\textwidth]{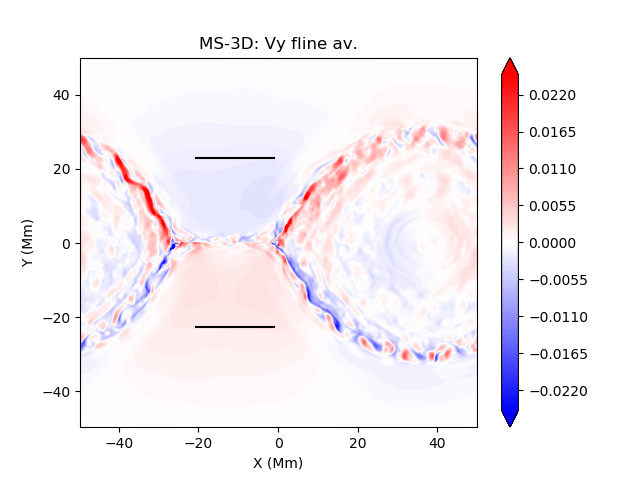}
    \includegraphics[width=0.33\textwidth]{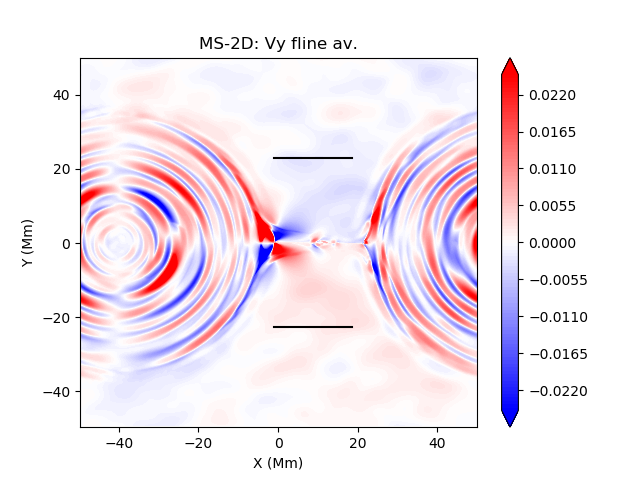} \\
    \includegraphics[width=0.33\textwidth]{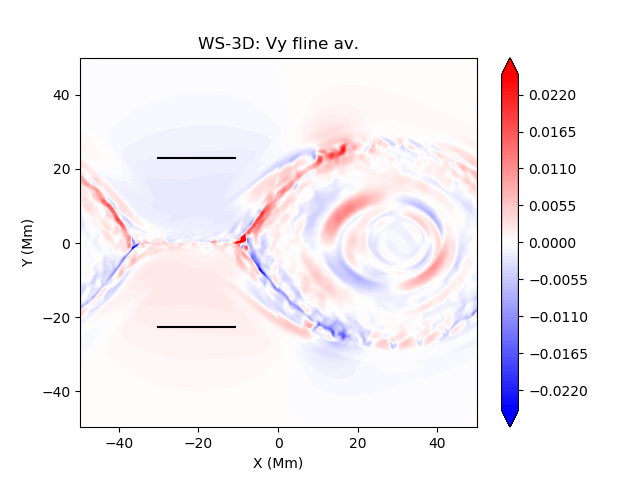}
    \includegraphics[width=0.33\textwidth]{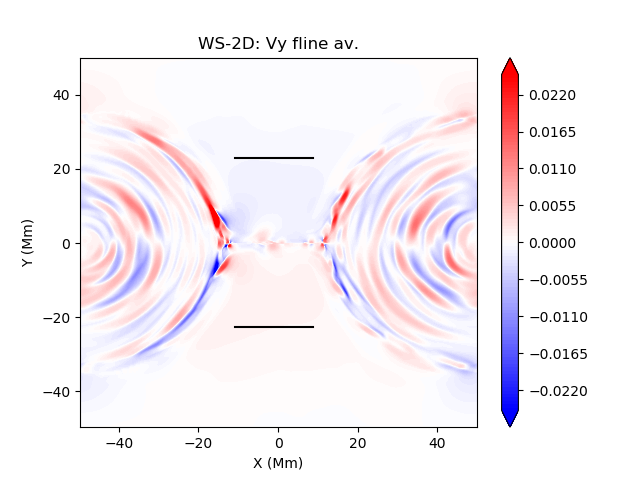}
    \caption{Velocity across the current sheet, $V_y$, averaged along fieldlines, in the $x,\ y\ (z=0)$ plane, for all 6 simulations. The snapshots are taken midway through the temporal windows defined in \S \ref{sec:results_initial}. In some cases the periodic data has been shifted in the x direction to place the primary current sheet near the center of the plot. 
    \label{fig:inflows}}
\end{figure}

\begin{figure}
    \centering
    \includegraphics[width=0.5\textwidth]{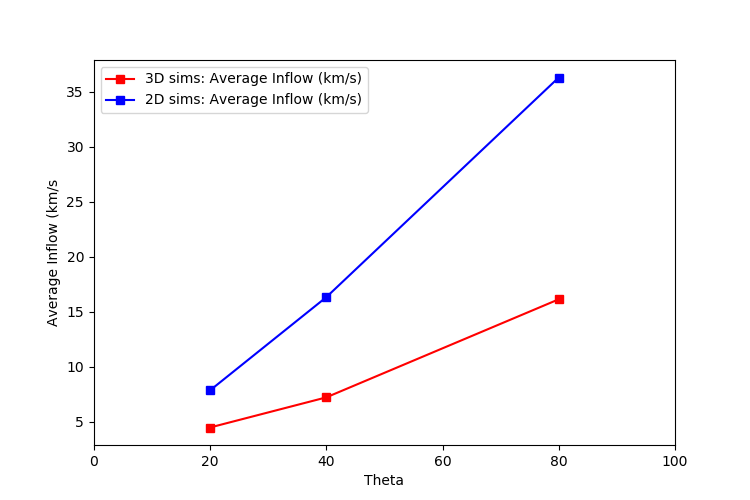}
    \caption{Average inflow for all 6 simulations.}
    \label{fig:inflow_only}
\end{figure}

The effect of the complex structure created by interacting oblique modes on the overall reconnection rate in the newly tearing current sheet can be evaluated by considering the average inflow to the newly formed current sheet. This average inflow is calculated as follows: The vertical flow is averaged along fieldlines (the integration along fieldlines in 2.5D is trivial), and shown in the $x,\ y$  plane at $z = 0$ in Figure \ref{fig:inflows} for each simulation. The time for each panel is taken to be when 80\% of the maximum change in internal energy has occurred (see Figure \ref{fig:internal_energy}). The inflow is averaged over a small $x$ range at a fixed $y$ index of 100 or 700, shown by the horizontal black lines in 
Figure \ref{fig:inflows}, and then averaged in time, where the temporal interval for each simulation is the interval between when 20\% and 80\% of the maximal internal energy change has occurred. Later in this section, \S \ref{sec:results_rates}, the normalized version of this inflow rate and its relationship with other normalized measures of reconnection rate is discussed. 

Figure \ref{fig:inflow_only} shows the results for all 6 simulations. In both 2.5D and 3D there is an increase in the average inflow velocity with increasing strength of reconnecting field (denoted by angle $\theta$), which is to be expected as the growth rates of the tearing instability increases with this field strength (see \citet{Baalrud_2012,2020ApJ...891...62L}, and Figure \ref{fig:theory_rates}. From Figure \ref{fig:inflow_only} it is clear that for a given reconnecting field value, the average inflow is faster in 2.5D than in 3D. This is a direct consequence of the disturbance of the current sheet plasmoid evolution by interacting oblique modes in 3D. 

It is also clear from Figure \ref{fig:inflow_only} that a linear fit of the average inflow to strength of the reconnecting field (or rotation angle $\theta$) 
 has a larger gradient for 2.5D than for 3D. It is here proposed that there are two major effects on the average inflow rate in 3D. The first has a positive effect and is the increased Alfv\'{e}n speed due to the large reconnecting field. The second has a negative effect and is the quantity and strength of interaction of the flux tubes formed by the oblique modes and hence the strength of the disturbance of the plasmoid evolution in the current sheet, which also increases with the reconnecting field strength. The resultant dependence of average inflow on reconnecting field strength thus has a reduced gradient in 3D when compared to 2.5D, where only the first effect has a role.

This also explains why the difference in average inflow between 2.5D and 3D is smallest at weak reconnecting field strength (low $\theta$), and largest at large strength (high $\theta$), as seen in Figure \ref{fig:inflow_only}, and is consistent with the prediction made in \S \ref{sec:results_initial} that the weak reconnecting field simulation would behave more like 2.5D due to its relatively strong guide field compared to reconneting field, and the absence of strongly growing oblique modes in the linear analysis.

%% file: results_rates.tex
\subsection{Normalized Reconnection Rates}
\label{sec:results_rates}

In the magnetic reconnection literature, there are multiple methods used to calculate a normalized magnetic reconnection rate, including inflows normalized by local Alfv\'{e}n speeds, reconnection electric fields or rate of change of reconnected flux, and heating rates \citep[see e.g.,][]{Comisso_2016}. So far, in this work, only the average inflow to the reconnection region has been discussed. In order to both contextualize the results, and ensure that the average inflow calculation described above is consistent with these normalized methods, three different normalized reconnection rates are calculated here.

\begin{table}[h]
\centering
\begin{tabular}{|c||c|c|c|c|} 
 \hline
 Name &   ${[V_{in}]}_{ave} (10^3 ~ m/s)$ &  $M_{1} = {[V_{in}/V_{A}]}_{ave} (10^{-3})$ & $M2 = \frac{dE/dt}{E_{free}/t_{A}} (10^{-3})  $ &  $M3 =\frac{d\phi/dt}{B_{x,0}v_{A,0}} (10^{-3}) $ \\ 
 \hline
\hline
 WS-3D &  4.61 & 8.08 & 8.47 & 6.54 \\
 \hline
 MS-3D &  7.57 & 5.62 & 6.94 & 4.69   \\
 \hline
 SS-3D &  16.13 & 4.80 & 5.95 & 3.25 \\
 \hline
 \hline
 WS-2D & 7.85 & 13.07 & 13.93 & 11.08 \\
 \hline
 MS-2D & 16.32 & 13.72 & 12.72 & 10.48  \\
 \hline
 SS-2D & 36.31 & 18.29 & 20.14 & 14.92 \\
 \hline
\end{tabular} 
 \caption{Summary of reconnection rates in all 6 simulations. \label{table:results}}
\end{table}

Table \ref{table:results} summarizes the different calculations of the three normalized magnetic reconnection rates (M1,M2,M3), though they are also visualized in later figures. Also shown is the averaged inflow discussed in \S \ref{sec:results_turbulence}.

\begin{figure}
    \centering
     \includegraphics[width=0.45\textwidth]{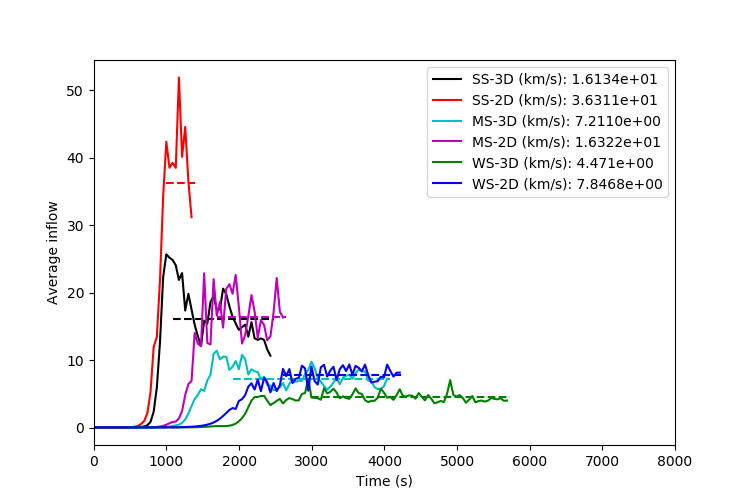} 
     \includegraphics[width=0.45\textwidth]{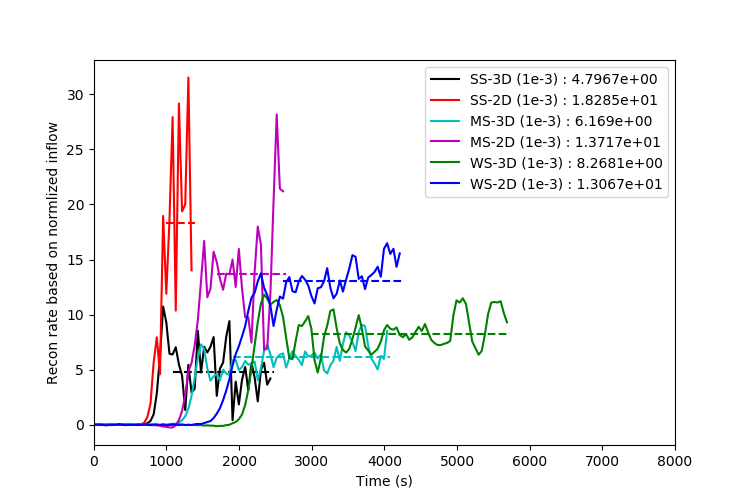}  \\
    \includegraphics[width=0.45\textwidth]{RATES_dEdt.png} 
    \includegraphics[width=0.45\textwidth]{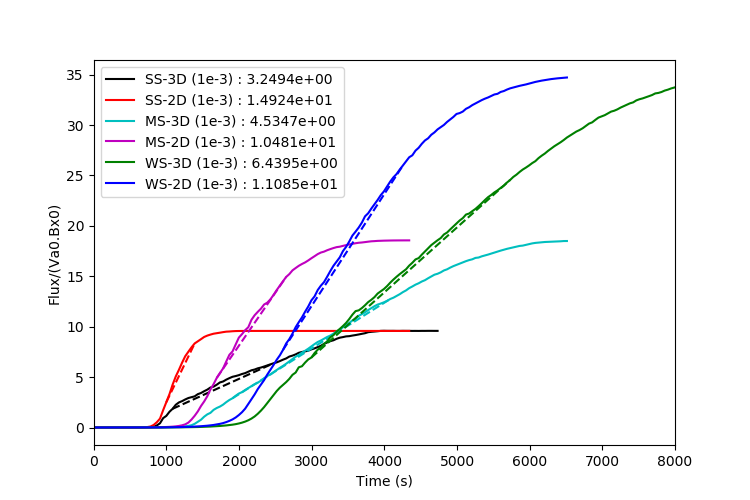} \\
    \includegraphics[width=0.45\textwidth]{RATES_inflow_only.png}
     \includegraphics[width=0.45\textwidth]{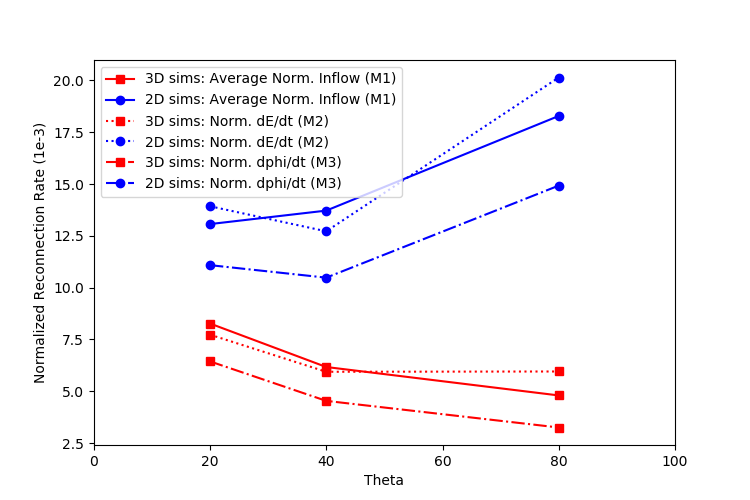}
    \caption{Top panels: Spatially averaged inflow (left) and normalized inflow (right) for all 6 simulations as a function of time. Middle panels: Time series of internal energy change (left) and reconnected flux (right) as a function of time. Bottom panels: Temporally and spatially averaged inflow (left) and normalized reconnection rate (right). The normalized reconnection rate is calculated three different ways in the bottom right panel (M1,M2,M3).
    \label{fig:recon_rates}}
\end{figure}

To remove some observer bias in all the calculations, and as discussed briefly in \S \ref{sec:results_initial}, the various reconnection rate calculations are made over a temporal interval which varies from one simulation to another. Figure \ref{fig:recon_rates}, middle left panel, shows the change in normalized internal energy for each of the 6 simulations. The window of calculation is defined as the interval at which the change in internal energy is 20\% and 80\% of the total change, and is represented by the dotted lines.

 The first normalized reconnection rate, denoted $M_1$ in Table \ref{table:results}, is based on the field-line integrated inflow normalized by the up-stream Alfv\'{e}n speed based on the reconnecting field component. It is calculated in a similar way to the average inflow described above, but rather than spatially and temporally averaging the vertical velocity, first it is normalized by Alfv\'{e}n speed. For a given $x$ location, the vertical velocity is divided by the maximum Alfv\'{e}n speed along a line in $y$ at that  $x$ location. Then the horizontal average is taken along the black horizontal lines shown in Figure \ref{fig:inflows}. The time series of the result for each simulation is shown in the top right panel of Figure \ref{fig:recon_rates}, along with the result for the un-normalized inflow in km/s (top left panel). This time series is then temporally averaged in the time window defined by the internal energy change, described above. This process can be represented by 
\begin{equation}
    M_{1} = {[{V_{in}(x)/\max{V_A(x)}]}_{ave,x}}_{ave,t}
\end{equation}
and is shown in the second column of Table \ref{table:results}.

The second reconnection rate, denoted $M_2$ in Table \ref{table:results}, is the average rate of change of the internal energy in the temporal window described above. As mentioned above, and shown in the middle left panel of Figure \ref{fig:recon_rates}, the change in internal energy is normalized by $E_{free}/t_{A}$  (where $t_A = L_{x}/V_{A,0}$) and so when the temporal derivative is taken, the result is non-dimensional and can be compared to $M_1$ above: 
\begin{equation}
    M_{2} = \frac{dE_{int}/dt}{E_{free}/t_A}
    \end{equation}

The third reconnection rate, denoted $M_{3}$ in Table \ref{table:results}, is based on the rate of change of reconnected flux. The flux, as in \citet{Huang_2016}, is calculated using the flux function along $y=0$, $z=0$ $A(x) = \int_{x_{min}}^{x}{B_{y}(x) dx}$, at any given time. This is normalized by $v_{A,0}B_{x,0}$ so that a temporal derivative will result in a dimension-less reconnection rate. This time series is shown for each simulation in the middle right panel of Figure \ref{fig:recon_rates}. The average rate of change is  calculated in the temporal window described above:
\begin{equation}
    M_{3} = \frac{\frac{d}{dt}\left[\max{A}-\min{A}\right]}{v_{A,0}B_{x,0}}.
\end{equation}

The three reconnection rates are plotted for all simulations in the bottom right panel of Figure \ref{fig:recon_rates}. Also shown in the bottom left panel is the average inflow calculation described in \S\ref{sec:results_turbulence}.

Recall that for the average inflow described in \S \ref{sec:results_turbulence}, the inflow for the 2.5D simulations has a monotonically increasing dependence on the shear angle, seen in the bottom left panel of Figure \ref{fig:recon_rates}. Since the guide field is fixed in this study, this corresponds to a monotonically increasing dependence on the reconnecting component of the field and the Alfv\'{e}n speed based on this component. This is the characteristic speed at which reconnected field lines move, so one might propose that normalizing this inflow by Alfv\'{e}n speed would result in no dependence of the resultant normalized rate on the shear angle, but there remains a small positive correlation for the average of the normalized inflow ($M_1$).

For the 3D simulations the average inflow (bottom left panel) also has a monotonically increasing dependence on shear angle, but with a lower gradient than in 2.5D, as discussed in \S\ref{sec:results_turbulence}. However, the normalized inflow rate $M_1$ has a monotonically decreasing dependence on the shear angle. This can be explained by recalling that there are two effects in the reconnection process in 3D: the reconnecting  field strength and the entangling produced by oblique modes, the latter of which increases with increasing reconnecting field. By normalizing the result by Alfv\'{e}n speed and thus removing some of the dependence on reconnecting field strength, the resultant plots primarily reflect the effect of the complex flows generated by oblique modes disrupting the plasmoid evolution, which increases with reconnecting field strength and has a negative effect on the reconnection process.

Comparing the two other methods of reconnection rate calculation ($M_2,M_3$) to the normalized inflow method ($M_1$), these general trends are reproduced, which provides some level of robustness to the results. However, note that for the medium shear angle 2.5D simulation (MS-2D) the flux and energy based rates results are lower than the low shear angle 2.5D simulation (WS-2D) thus breaking the monotonically increasing dependence seen in $M_1$. This is also reflected in plateaus  around 2200 s in the flux and internal energy plots (magenta lines) in Figure \ref{fig:recon_rates}, middle panels. In this particular simulation, the temporary pause in coalescence of plasmoids momentarily slows down both the conversion of flux and the resultant expulsion of flux and the following viscous heating. These plateaus cause the average reconnection rates, calculated using energy and flux, to be lower than the rate based on inflows.

The difference seen here between the normalized reconnection rate for 2.5D and 3D is in general agreement with the results of \citet{Huang_2016}. Here the flux-based reconnection rates are [0.00325,0.01492] for 2.5D and 3D, respectively, for the strong reconnecting field case (SS) where the ratio $B_{x,0}/B_{z,0} \approx 0.7$. In \citet{Huang_2016}, which has a slightly stronger ratio of reconnecting field to guide field, $B_{x,0}/B_{z,0} \approx 1$, the reconnection rates in the early stages are similar to those here, despite the initial setup being vastly different in the simulations of \citet{Huang_2016} and those that are presented here.

{\cite{2020ApJ...894L...7Z} 
have shown that, because of compressibility effects, the reconnection rate increases by nearly a factor of 2 as the inflow beta decreases from larger than to smaller than unity. The inflow beta varies from 0.002 to 0.003 in our simulations with different shear because the guide field and pressure are constant. This has negligible influence on the reconnection rate because it is far into the compressible regime.}

{A spatial oscillation along y can be seen in the inflow region of case SS-2D at the top right of Figure 7. A corresponding temporal oscillation is present at the top right of Figure 9 (red curve). Some of this is a remnant of the initial perturbation, and some of it is associated with the coalescence of plasmoids, which generates waves propagating outward from the sheet. We do not believe this has a significant effect on the temporally averaged inflow velocity - which is represented by the horizontal line segment - but this cannot be ruled out.}

%% file: discussion.tex
\section{Discussion}
\label{sec:discussion}

Through a limited parameter study of magnetic reconnection in 2D and 3D current sheets, varying the strength of the reconnecting field, the effect of oblique modes in 3D on efficient tearing in a current sheet was investigated. All simulations evolved in generally the same way, with the initial condition creating a large flux rope with a newly formed current sheet, which is then unstable to further tearing. The global evolution of this new current sheet was studied across the parameter study, using multiple measure of reconnection rate.

 In 3D current sheets with a guide field, the tearing modes are flux tubes that are  aligned with the local magnetic field direction and therefore have different angles of obliquity as a function of distance from the center of the sheet. The resultant complex structure generated by their non-linear 
interaction, restricts the efficient 2D plasmoid evolution that would occur in their absence. In the 3D simulations the oblique modes interfere with the newly forming primary current sheet that forms together with the main flux rope. 
The otherwise organized behavior seen in 2D – formation of islands and ejection to the ends of the sheet – becomes messy, with tangled flux tubes. This clogs up the flow. The effect is greater for large reconnecting field strength, because more oblique modes are present in the system. 

Thus, for all values of reconnecting field strength, the 3D reconnection is slower than in 2D, based on rate calculations based on inflows, flux changes, and internal energy changes (heating).
As the reconnecting field strength increases in this parameter study, the strength of the oblique modes increases and so the 3D simulations show a weaker normalized reconnection rate, though the non-normalized average inflow still increases with shear angle, as the positive effect of the reconnecting field strength on the reconnection rate  via the Alfv\'{e}n speed is still present in the calculation. In addition, the low shear angle 3D simulation is closer in rate to the equivalent 2D simulation, as the weak reconnecting field system has very weak oblique modes, and is much more 2D-like, as the guide field is effectively stronger at weaker reconnecting field strength.

One caveat related to the numerical solution of the governing equations is that the minimum resolution of 0.0125 Mm is clearly not enough to fully resolve the detailed evolution of the plasmoids in the later stages, which is why a detailed analysis of the scaling of the size and frequency, as done in \citet{Huang_2010} has not been performed. As mentioned in \S\ref{sec:numerics}, the early evolution is well captured, which includes the disruption of the original current sheet and the formation of the new flux rope and primary current sheet.
Therefore, while it is reasonable to assume that the onset of the breakup of the newly formed current sheet, and hence the reconnection rate in the early stages of this breakup is captured with some fidelity by these simulations, the same cannot be said for the later stages. 

The conclusion of complex-flows, driven by oblique modes, interfering with the efficient magnetic reconnection process in current sheets points to the larger question of how turbulence or noise interact with current sheets that are undergoing tearing. Near uniform or white noise as used as the seed in our initial current sheet will give more or less the same preference to all modes.  Color noise as is present in the oblique-dominated 3D simulation, will give clear positive or negative preference to various modes in the current sheet evolution. 

\citet{Huang_2016} studied the self-sustained turbulence generated by oblique modes in a reconnecting current sheet that rapidly thins from its initial configuration. In this study, initially the 3D simulations are slower than 2D as the 2D simulations generate lots of parallel modes (plasmoids), but then the 2D rate decreases as these plasmoids coalescence, while the 3D rate slowly increases, until the two rates are comparable. In our parameter study, the 2D and 3D rates do not converge. It is proposed that the primary reason for this apparent discrepancy is that the geometry of the fragmenting current sheet is different in \citet{Huang_2016} compared to this study. The current sheet is in initial non-equilibrium in \citet{Huang_2016} and undergoes rapid thinning, and as it tears, the two flux-rope system used to create a current sheet means that reconnection continues to drive the thinning (newly formed field lines surround both flux ropes and pull them together). This will cause an overall increase of reconnection rate with time. Our simulations show that the inflow is complex in 3D even when the sheet is not driven.

Note that this study and that of \citet{Huang_2016} are different from the turbulence-driven reconnection studies discussed in \citet[e.g.,][]{Kowal_2009, Lazarian_2015}. Those systems are inherently turbulent. The turbulence is driven by processes other than reconnection; it is not generated by the reconnection itself. We do not dispute the claim that Sweet-Parker type reconnection is faster in such turbulent systems than in undisturbed systems. However, our current sheets are thicker than the Sweet-Parker scale. {Furthermore, we agree with Klimchuk and Antiochos (2021) that the low-beta, line-tied, magnetically closed corona is not inherently turbulent. It evolves primarily quasi-statically, punctuated by many localized, short-lived bursts of turbulent activity associated with magnetic reconnection at a myriad of small current sheets. The Fourier power spectra of magnetic energy in some simulations that are often attributed to turbulent fluctuations are instead an indication of the abrupt magnetic field rotations in the quasi-static current sheets.} 

\begin{figure}
    \center
    \includegraphics[width=1.0\textwidth]{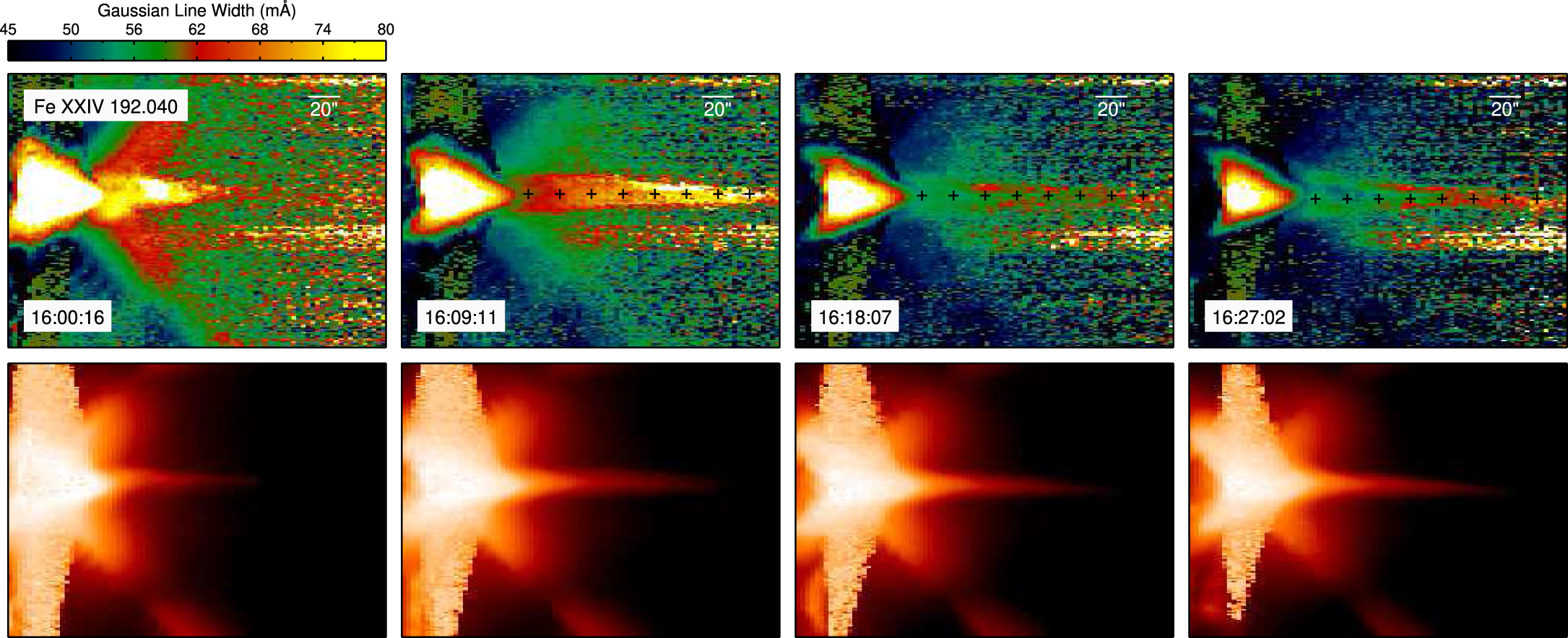}
 \caption{Figure is taken from \cite{Warren_2018}  investigating an X8.3 flare at the west limb of the sun on 2017 September 10. Line widths and intensities derived from the EIS Fe XXIV 192.04 \AA \ line. 
 The broadest profiles tend to occur early in the event and broadening appears to increase with height above the arcade. Line widths also decrease with time during the
event. (\copyright AAS. Reproduced with permission)}
\label{fig:flare}
\end{figure}

While some plasmoid-like structures have been observed in coronal current sheets \citep{Pankaj_2019}, the lack of ubiquitous observations of plasmoids may point to the entangling/turbulence created by interacting oblique modes in 3D. This effect may also vary temporally, as the effective guide field changes in time in evolving current sheets. {Such is the case with eruptive flares, for example. The flare occurs at a vertical current sheet that forms when a magnetic arcade is stretched out by the eruption. It is the shear component of the pre-event arcade that becomes the guide field of this newly formed current sheet. Since arcades typically have decreasing shear with distance from the polarity inversion line, the guide field decreases with time as the flare reconnection proceeds \citep{Qiu_2017,Dahlin2022}}. {This may help address observations such as those reproduced from \citet{Warren_2018} in Figure \ref{fig:flare}, which shows the temporal evolution of the measured intensities and widths of an Fe XXIV line in the vicinity of an eruptive flare current sheet.} In these observations, the line widths are larger earlier in the flare, indicating that turbulence may be stronger early on, and that a decrease in guide field with time might explain a decrease in oblique-dominated evolution and hence a decrease in turbulent line broadening.  Turbulent reconnection has been inferred from recent observations \citep{Cheng_2018,Polito_2018}, and turbulence has also been suggested as an explanation for the confinement of bremsstrahlung-producing electrons in the flares, which can enhance the angular scattering rate as well as reducing the thermal electron transport, see \cite{2019ApJ...880...80B,2021SoPh..296..147K} and references wherein.

It is important to distinguish between force-free and non-force-free current sheets. In force-free sheets, like the ones we have simulated, the total field strength is constant, so the guide field is stronger in the center of the sheet than outside. In non-force-free sheets, the guide field is constant, and gas pressure counterbalances the reduced magnetic pressure at the center. Neutral sheets have no guide field, by definition, so even though their anti-parallel fields might be thought of as having a 180 degree shear angle, oblique modes do not exist, because a guide field is required. The shear angle of force-free current sheets is given by the external values of the reconnecting and guide field components. Larger reconnecting field is associated with more oblique modes.

It is also worth noting that the effect of the relative strength of guide field and reconnecting field has been investigated in 2D and 3D in the kinetic regime \citep{Goodbred_2021}. In that work, there was a slightly higher  energy conversion rate in 2D compared to 3D, and the presence of oblique modes in 3D was important.